\documentclass[fleqn,usenatbib]{mnras}

\usepackage{newtxtext,newtxmath}

\usepackage[T1]{fontenc}
\usepackage{ae,aecompl}

\usepackage{graphicx}	
\usepackage{amsmath}	
\usepackage{amssymb}	
\usepackage{adjustbox}
\usepackage[justification=centering]{caption}
\usepackage[]{algorithm2e}

\title[Searching for Exoplanets using AI]{Searching for Exoplanets using Artificial Intelligence}

\author[K. A. Pearson et al.]{
Kyle A. Pearson,\thanks{E-mail: pearsonk@lpl.arizona.edu }
Leon Palafox,
and Caitlin A. Griffith
\\
Lunar and Planetary Laboratory, University of Arizona, 1629 East University Boulevard, Tucson, AZ, 85721, USA\\
}

\date{Accepted XXX. Received YYY; in original form ZZZ}

\pubyear{2017}

\begin{document}
\label{firstpage}
\pagerange{\pageref{firstpage}--\pageref{lastpage}}
\maketitle

\begin{abstract}
In the last decade, over a million stars were monitored to detect transiting planets. Manual interpretation of potential exoplanet candidates is labor intensive and subject to human error, the results of which are difficult to quantify. Here we present a new method of detecting exoplanet candidates in large planetary search projects which, unlike current methods uses a neural network. Neural networks, also called ``deep learning'' or ``deep nets'' are designed to give a computer perception into a specific problem by training it to recognize patterns. Unlike past transit detection algorithms deep nets learn to recognize planet features instead of relying on hand-coded metrics that humans perceive as the most representative. Our convolutional neural network is capable of detecting Earth-like exoplanets in noisy time-series data with a greater accuracy than a least-squares method. Deep nets are highly generalizable allowing data to be evaluated from different time series after interpolation without compromising performance. As validated by our deep net analysis of Kepler light curves, we detect periodic transits consistent with the true period without any model fitting. Our study indicates that machine learning will facilitate the characterization of exoplanets in future analysis of large astronomy data sets. 
\end{abstract}




\section{Introduction}

\begin{figure}
\centering
\hspace*{-.45in}
\includegraphics[scale=0.65]{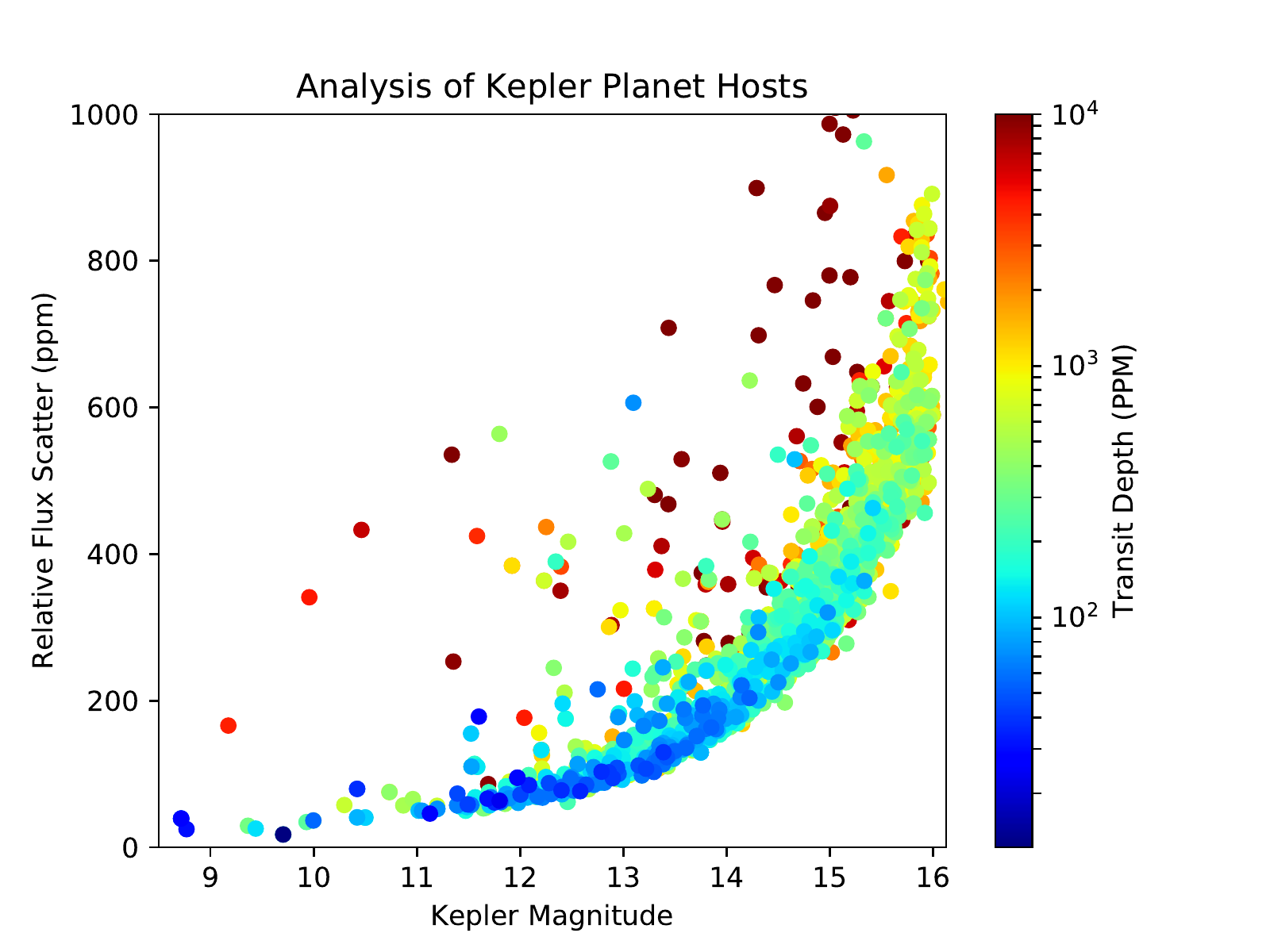}
\caption[Time series]{ The precision of Kepler light curves are calculated for all planet-hosting stars using the 3rd quarter of Kepler data. The scatter in the data is calculated as the average standard deviation from 10 hour bins, and averaging over small time windows minimizes the large scale variations caused by stellar variability. The transit depths of small planets are currently pushing the limits of Kepler and in some cases the depths are below the noise scatter. }
\label{KeplerMag}
\end{figure}

Transiting exoplanets provide a remarkable opportunity to detect planetary atmospheres through spectroscopic features. During primary transit, when a planet passes in front of its host star, the light that transmits through the planet's atmosphere reveals absorption features from atomic and molecular species. Currently 3,513 exoplanets have been discovered from space missions (Kepler \citep{Borucki2010}, K2 \citep{Howell2014} and CoRoT \citep{Auvergne2009}) and from the ground (HAT/HATnet \citep{Bakos2004}, SuperWASP \citep{Pollacco2006}, KELT \citep{Pepper2007} ). Future planet hunting surveys like TESS, PLATO and LSST plan to increase the thresholds that limit current photometric surveys by sampling brighter stars at faster cadences and over larger field of views (\citealt{LSST2009}; \citealt{Ricker2014}; \citealt{Rauer2014}). Kepler's initial four-year survey revealed $\sim$15$\%$ of solar type stars have a 1--2 Earth-radius planet with an orbital period between 5--50 days (\citealt{Fressin2013}; \citealt{Petigura2013}). The detection of such small Earth-sized planets are difficult because the transit depth, $\sim$100 ppm for a solar type star, reaches the limit of current photometric surveys and is below the average stellar variability (see Figure \ref{KeplerMag}). Stellar variability is present in over 25$\%$ of the 133,030 main sequence Kepler stars and ranges between $\sim$950 ppm (5th percentile) and $\sim$22,700 ppm (95th percentile) with periodicity between 0.2 and 70 days \citep{Mcquillan2014}. The analysis of data in the future needs to be both sensitive to Earth-like planets and robust to stellar variability.


\begin{figure}
\centering
\includegraphics[scale=0.65]{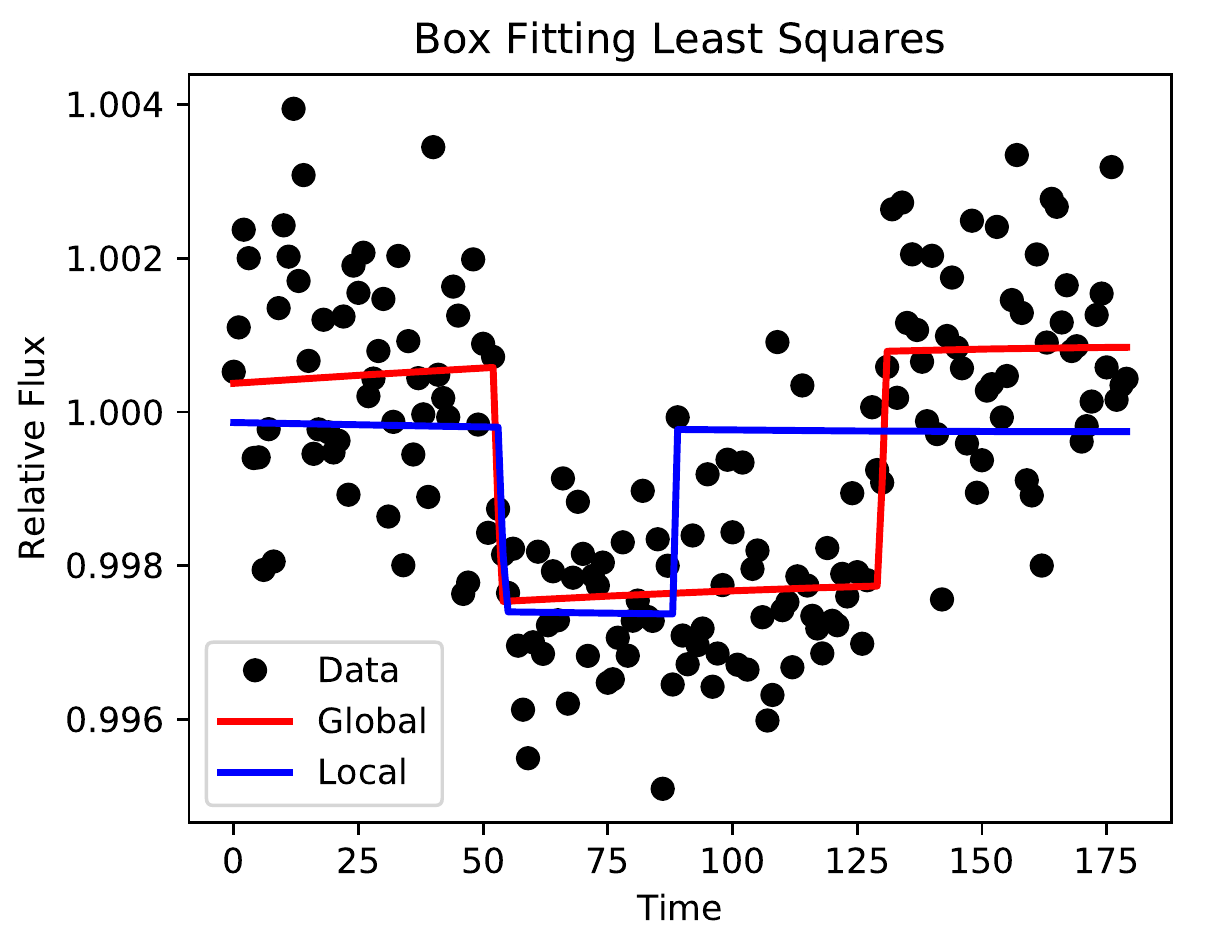}
\caption[BLS]{ Past transit detection algorithms are accomplished by correlating the data with a simple box model through a least-squares optimization. The global solution is shown in red and was initialized with the parameters used to generate the data. The blue line uses randomly initialized parameters and shows a local solution. Least Squares algorithms are susceptible to finding local minima in their parameter space, which can lead to false detections of transits. The data and model were generated with a box transit function plus a sinusoidal systematic trend to simulate variability. }
\label{BLS}
\end{figure}

Common techniques to find planets maximize the correlation between data and a simple transit model via a least-squares optimization, grid-search, or matched filter approach (\citealt{Kovacs2002}; \citealt{Jenkins2002}; \citealt{Carpano2003}; \citealt{Petigura2013}). A least-squares optimization aims to minimize the mean-squared error (MSE) between data and a model. Since the transit parameters are unknown a priori, a simplified transit model is constructed with a box function. Least-square optimizers are susceptible to finding local minima (see Figure \ref{BLS}) when trying to minimize the MSE and, thus, can result in inaccurate transit detections unless the global solution can be found. When individual transit depths are below the scatter, as is the case for Earth-like planets currently, constructively binning the data can increase the signal-to-noise (SNR). Grid-searches utilize binning by performing a brute-force evaluation over different periods, epochs and durations to search for transits either with a Least-squares approach \citep{Kovacs2002}; or matched-filter \citep{Petigura2013}. A matched filter approach tries to optimize the signal of a transit by convolving the data with a hand-designed kernel/filter to accentuate the transit features.

The SNR of a transit detection can be maximized when convolving data with the optimal filter. However, solving for the optimal filter in the case of varying transit shapes cannot be done analytically, so kernels are hand-made to approximate what the human user thinks is best. Deep learning with convolutional neural networks (CNN) have previously been used to solve similar kernel optimization problems \citep{Krizhevsky2012}. Recently, these CNNs have been able to beat human perception at object recognition (\citealt{He2015}; \citealt{Ioffe2015}). Other heuristics at automating planet finding have been developed using Random Forests (\citealt{McCauliff2015}; \citealt{Mislis2016}), Self-Organizing Maps \citep{Armstrong2016}, and k-nearest neighbors \citep{Thompson2015}. These algorithms are used to remove a substantial fraction of false positive signals prior to detecting the planet's signal. The fraction of false positives can be reduced by requiring at least three self-consistent transit detections instead of a single one (\citealt{Petigura2013}; \citealt{Burke2014}; \citealt{Rowe2015}; \citealt{Coughlin2016}).

Complications can arise if the input data is not quiescent where the instrumental systematics or stellar variability are marginal compared to the signal of the transit and change with time. De-trending the data from these effects requires a prior information, where an assumption is made about the behavior of the effects perturbing the data. Techniques to de-trend light curves from instrumental or stellar systematics have been successful in the past using Gaussian Processes (e.g. \citealt{Gibson2012}; \citealt{Aigrain2015}; \citealt{Crossfield2015}), PCA (e.g. \citealt{Zellem2014}), ICA (e.g. \citealt{Waldmann2013}; \citealt{Morello2015}), wavelet based approaches (e.g. \citealt{Carter2009}), or more classical aperture photometry de-trending (e.g. \citealt{Armstrong2014}). Typically these methods involve modeling the systematic trend and subtracting it from the data before characterizing the exoplanet signal. This disjointed approach may allow the transit signal to be partially absorbed into the best-fit stellar variability or instrument model. In this way, the methods may be over-fitting the data and making each transit event appear shallower. \cite{Foreman-Mackey2015} propose an alternate technique that models the transit signal simultaneously with the systematics and find a large improvement on planet detection but at the cost of computational efficiency.

The ideal algorithm for detecting planets should be fast, robust to noise and capable of learning and abstracting highly non-linear systems. A neural network trained to recognize planets with simulated data provides the ideal platform. Deep learning with a neural network is a computational approach at modeling the biological way a brain solves problems using collections of neural units linked together (\citealt{Rosenblatt1958}; \citealt{Newell1969};). Deep nets are composed of layers of ``neurons'', each of which are associated with different weights to indicate the importance of one input parameter compared to another. Our neural network is designed to make decisions, such as whether or not an observation detects a planet, based on a set of input parameters that treat, e.g. the shape and depths of a light curve, the noise and systematic error, such as star spots. The discriminative nature of our deep net can only make a qualitative assessment of the candidate signal by indicating the likelihood of finding a transit within a subset of the time series. Within a probabilistic framework, this is done by modeling the conditional probability distribution $P(y|x)$ which can be used for predicting $y$ (planet or not) from $x$ (photometric data). The advantage of a deep net is that it can be trained to identify very subtle features in large data sets. This learning capability is accomplished by algorithms that optimize the weights in such a way as to minimize the difference between the output of the deep net and the expected value from the training data. Deep nets have the ability to model complex non-linear relationships that may not be derivable analytically. The network does not rely on hand designed metrics to search for planets, instead it will \textit{learn} the optimal features necessary to detect a transit signal from our training data.

Neural networks have been used in a few planetary science applications including multi-planet prediction and atmospheric classification. \cite{Kipping2017} train a neural network to predict multi-planet systems. While this neural network does not focus on directly detecting transits, it models the correlation between orbital period, planetary radius and mass to predict the presence of additional bodies in a planetary system. Data from TESS is limited to short period planets (i.e. $P$ < 13.7 days) but predictions about longer period planets can be modeled from correlations in Kepler data. Neural networks provide an effective tool to model such non-linear systems when no analytic formula is known or even possibly derivable. Another application of deep learning in exoplanet science predicts atmospheric compositions based on emission spectra \citep{Waldmann2016}. Additionally, \citealt{George2017} introduces a new method for time-series classification and regression of highly noisy gravitational waves using raw time-series inputs in a 1D convolutional neural network.

In this paper we design various deep learning algorithms to recognize planetary transit features from a training data set. In Section 3, we explain the architecture of our deep learning algorithms. In Section 4, we explore the sensitivities of each algorithm at detecting planets in noisy data. Section 5 covers the time series evaluation of multiple transits and validates the detection algorithm against known planets in Kepler data. In section 6 and 7, we discuss our results and summarize our findings in the conclusion.

\section{Implementation and Training}

\begin{table*}
  \caption{Summary of Parameters}
    \begin{tabular}{lcrr}
  Training Parameters & & Values & Unit \\
  \hline &&& \\ 
  Transit Depth &  R$_{p}^{2}$/R$_{s}^{2}$ & 200,500,1000,2500,5000,10000 & ppm \\
  Orbital Period & P & 2,2.5,3,3.5,4 & days\\
  Inclination & i & 86,87,90 & degrees \\
  Noise Parameter & $\sigma_{tol}$ & 1.5,2.5,3.5,4.5 &  \\
  Phase Offset & $\phi$ & 0, $\pi$/3, 2$\pi$/3, $\pi$ & \\
  Wave Amplitude & A & 250,500,1000,2000 & ppm \\
  Wave Period & $\omega$ & 6./24, 12./24, 24./24 & days \\
 Amplitude Variability Period & $P_A$ & -1, 1, 100 & days \\
 Wave Variability Period & $P_\omega$ & -3, 1, 100 & days \\
  &&& \\  &&& \\

  Sensitivity Test & & Values & Unit \\
  \hline &&& \\
  Noise Parameter& $\sigma_{tol}$ & 0.25,0.5,0.75,1,1.25,1.5,1.75,2.25,2.5,2.75,3 &  \\
  &&& \\  &&& \\


  Fixed Transit Parameters && \\
  \hline && \\
  Scaled Semi-major Axis & a/R$_{s}$ & 12 & \\
  Eccentricity & $e$ & 0 & \\
  Arg. of Periastron & $\Omega$ & 0 & \\
  Linear Limb Darkening & $u_{1}$ & 0.5 & \\
  Mid Transit Time & $t_{mid}$ & 1 & \\
  Window Size & & 360 & minutes \\
  Cadence & dt & 2 & minutes \\
  Minimum Time & $t_{min}$ & 0.875 & days \\
  Maximum Time & $t_{max}$ & 1.125 & days \\
	&&& \\
  Training Samples & 311040 && \\
  Test Samples & 933120 &&\\
  \end{tabular}
 \label{tab:training}
\end{table*}

\begin{figure*}
\centering
\hspace*{-.125in}
\includegraphics[scale=0.5]{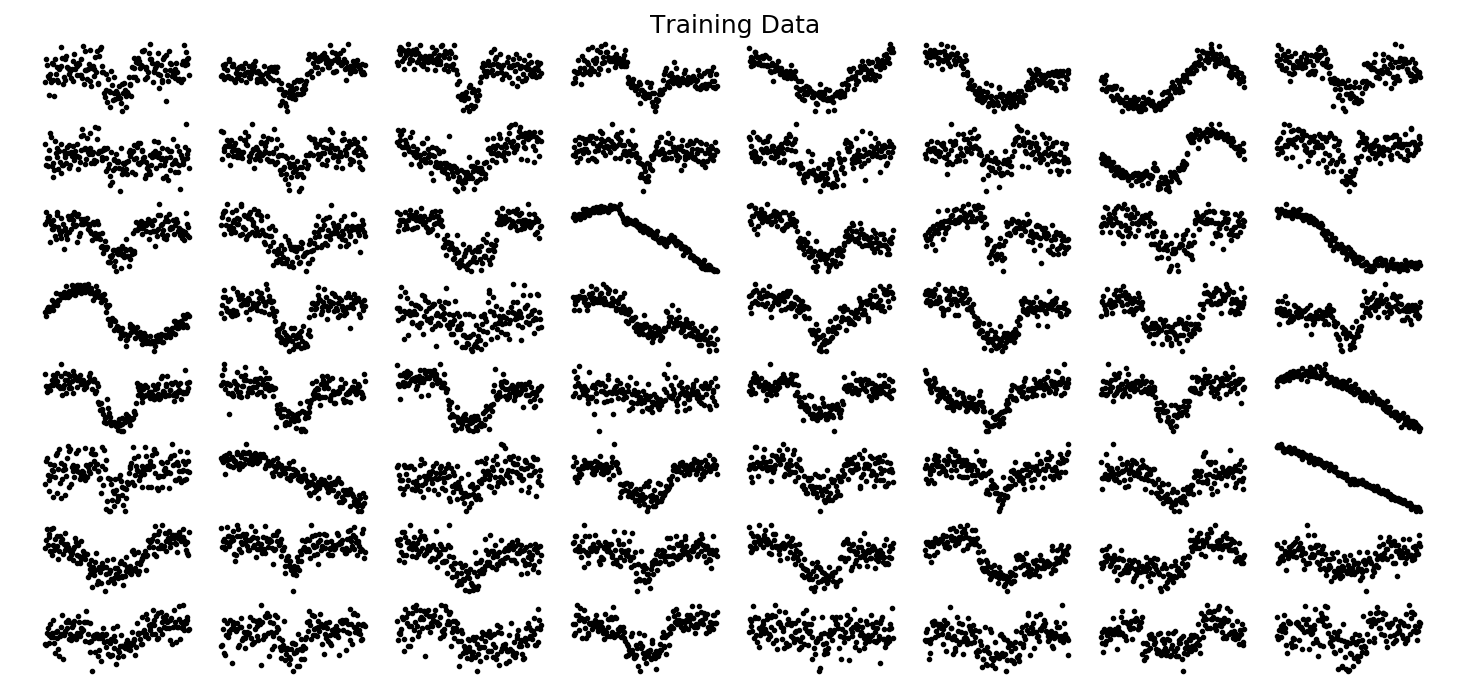}
\caption[Training Data]{ A random sample of our training data showing the differences between light curves and systematic trends. Each transit was calculated with a 2 minute cadence over a 6 hour window and the transit parameters vary based on the grid in Table \ref{tab:training}. }
\label{training_data}
\end{figure*}

Simulated training data is used to teach our deep nets how to predict single planetary transits in noisy photometric data. The simulated data is similar to what we would expect from a real planetary search survey. After the deep nets are trained, we use the network to assess the likelihood of potential planetary signals in data it has not seen before.

\begin{figure*}
\centering
\hspace*{-.125in}
\includegraphics[scale=0.8]{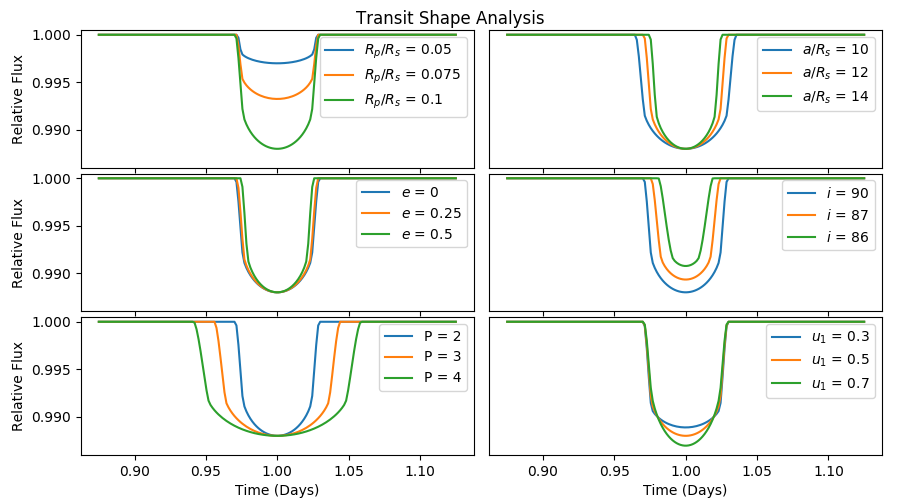}
\includegraphics[scale=0.8]{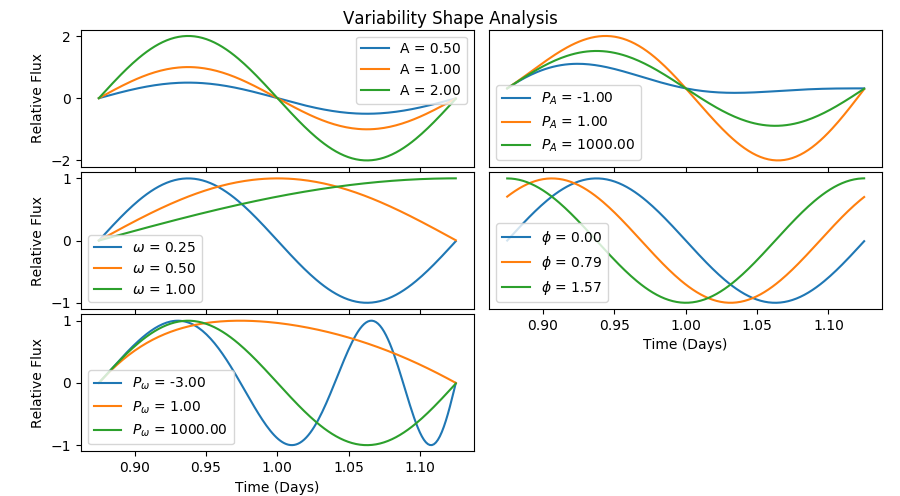}
\caption[Shape Analysis]{ The two plots show how each parameter can change the shape of a light curve as a function of the physical characteristics of the transit (top panels) and systematic effects due to, e.g., star spots (bottom panels). We generate training samples using parameters that greatly vary the overall shape (see Table 1). The default transit parameters are varied one by one with the initial parameter set; $R_p/R_s$=0.1, $a/R_s$=12, $P$=2 days, $u_1$=0.5, $u_2$=0, $e$=0, $\Omega$=0, $t_mid$=1. We neglect mid transit time, eccentricity, scale semi-major axis and limb darkening as varying parameters in our training data set. Our formulation for quasiperiodic variability is shown in Equation 1. The default stellar variability parameters are $A$=1, $P_A$=1000, $\omega$=6./24, $P_\omega$=1000, $\phi$=0. Each parameter is varied one by one from the default set and then the results are plotted. }
\label{shape_analysis}
\end{figure*}

\subsection{Training Data Set}

We generate a total of 311040 transits and non-transit data to train our deep nets. The training data are computed from a discretely sampled 9 dimensional hypergrid in our parameter space (see Table \ref{tab:training}). The parameters limit transit duration to 30 minutes and no longer than 4 hours, which spans 1./12 to 2./3 of the time domain. We chose our parameters to mimic data from real search surveys by encompassing many possible systematic shapes and transit sizes. The equation used to generate our noisy data with a quasi-periodic systematic trend is
$$
t' = t - t_{min}
$$
$$
A(t') = A + A \sin\left(\frac{2\pi t'}{P_{A}}\right)
$$
$$
\omega(t') = \omega + \omega \sin\left(\frac{2\pi t'}{P_{\omega}}\right)
$$
\begin{equation} \label{data}
F_{transit}(t) * \mathcal{N}\left( \frac{R_{p}^{2}}{R_{s}^{2}} / \sigma_{tol} \right) * \left(1+A(t')\sin\left(\frac{2\pi t'}{\omega(t')} + \phi\right)\right)
\end{equation}

\noindent where $F_{transit}(t)$ is the transit function given by \citealt{Mandel2002}, $A$ is the amplitude of our simulated stellar variability, $\omega$ is the period of oscillation, $\phi$ is the phase shift, and $\mathcal{N}$ is a Gaussian distribution used to generate random numbers with a mean of 1 and standard deviation of ($R_{p}^{2}$/R$_{s}^{2}$)/ $\sigma_{tol}$ and $R_{p}^{2}$/R$_{s}^{2}$ is the normalized radius ratio between the planet and star. Non-transit noisy data is generated in a similar manner, but without the transit signal, $F_{transit}(t)$.

The simulated data has a quasiperiodic systematic trend, like signals found in Kepler data \citep{Aigrain2015}. The simulated variability represents a sinusoid with varying frequency and amplitude given our choice of parameters, $P_A$ and $P_\omega$. We generate a standard sinusoidal model with no variation in amplitude or frequency when $P_A$ and $P_\omega$ are equal to 1000 days. The amplitude of the sinusoidal function will double over the span of 6 hours when $P_A$ is equal to 1 day. The amplitude of the sinusoidal function will diminish to zero when $P_A$ is equal to -1 day. It is physically unrealistic to have a negative period, however we use it as a mathematical tool for achieving the shape we want. The oscillation period will double when $P_\omega$=1 day and will diminish by half when $P_\omega$=-3. A shape analysis for each varying parameter in our systematic trend is shown in Figure 14.

\begin{equation} \label{noise}
 \mathcal{N}\left( \frac{R_{p}^{2}}{R_{s}^{2}} / \sigma_{tol} \right) * \left(1+A(t')\sin\left(\frac{2\pi t'}{\omega(t')} + \phi\right)\right)
\end{equation}

Each transit light curve has a non-transit sample using the same systematic parameters but newly generated noise of the same distribution shape and size. This allows our deep net to differentiate between transit and non-transit. The synthetic data are normalized to unit variance and have the mean subtracted off prior to input in the deep nets. Various light curves and systematic trends are shown in Figure \ref{training_data}. The variable light curve shapes and systematic trends are shown in appendix Figure \ref{shape_analysis}. 

\begin{figure*}
\centering
\includegraphics[scale=0.65]{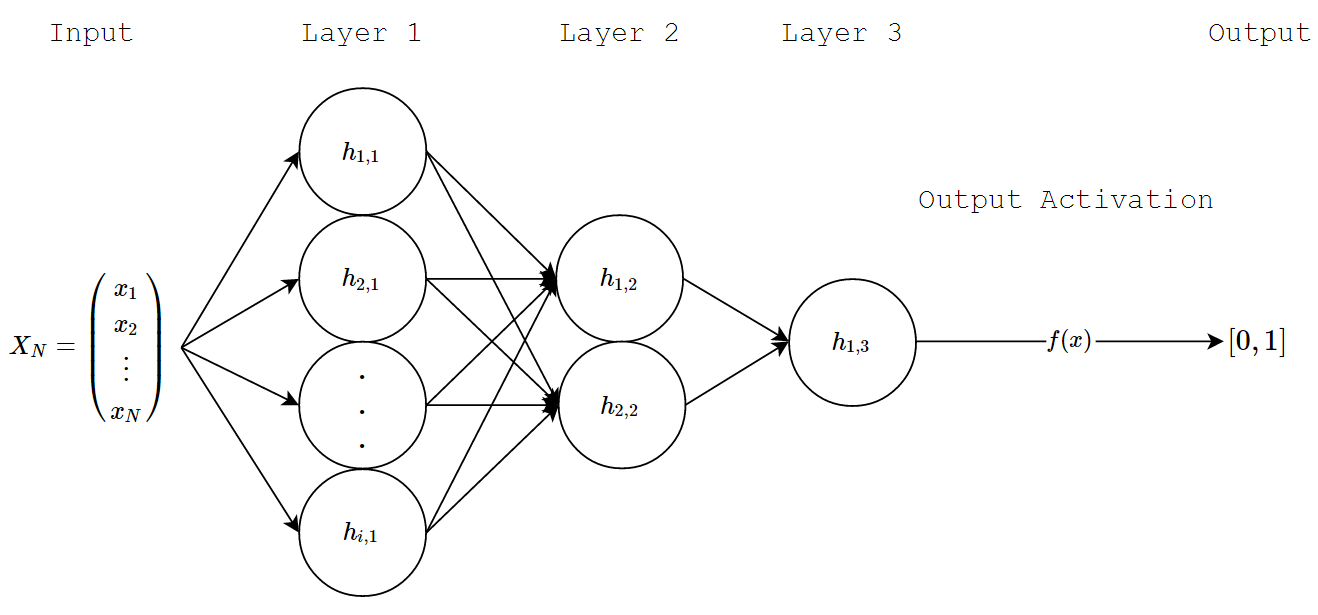}
\caption[Neural Network]{ The architecture for a general neural network is shown. Each neuron in the network performs a non-linear transformation on the input data to abstract the information for deeper layers. The output of each neuron is modified with our activation function (ReLU) and then used as the input for the next layer. The final activation function of our network is a sigmoid function that outputs a probability between 0 and 1, suggesting the signature of a planet is present (or not). }
\label{NN}
\end{figure*}


\subsection{Neural Network Architecture}
 
An artificial neural network is a computational approach at modeling the biological way a brain solves problems. It uses a collection of neural units connected to many others and has links/synapses which can be enforced or inhibited through the activation state. Networks are composed of layers of ``neurons'' (see Figure \ref{NN}), generally referred to as Restricted Boltzmann Machines (RBM), each of which has a set of input parameters, each of which are associated with a weight to indicate the importance of one input parameter compared to another. The circles in the graph represent one unit or neuron in the network and $h$ represents a transformation the neuron performs on the input data to abstract the information for the next layer. A neuron also has a bias, which acts like a threshold number for the final decision of that neuron, whether a yes/no decision or a probability. A fully connected neural network is one where each neuron in a higher layer uses the output from every neuron in the previous layer as the input. We implement a fully connected multi-layer perceptron for each of our deep nets.

For example, a perceptron is a type of neuron that takes several inputs, $x_1$, $x_2$,..., (e.g. photometric data) and produces a single output. Weights, $w_1$, $w_2$,..., are introduced to express the importance of the respective inputs to the output. The neuron's output is determined by whether the weighted sum $\sum_jw_j x_j$
is less than or greater than some threshold value, which is quantified by the bias, $b$. Just like the weights, the threshold is a real number, which is a parameter of the neuron. To put it in more precise algebraic terms:

$$
{\rm Output} = 0 ~~~~~ {\rm if} ~~\sum_jw_j x_j + b ~~\le ~~0 
$$

$$
{\rm Output} = \sum_jw_j x_j + b ~~~~~ {\rm if}~~ \sum_jw_j x_j + b ~~> ~~0
$$

\noindent Using simplified notation, $\sum_jw_j x_j$ = $\bf{w} \cdot \bf{X}$. The neurons transform the input data with a weighted sum and then use a non-linear function as the ``activation'' that defines the neuron output. The activation function we use is the rectified linear unit (reLU) and it zeros the output of a neuron if it has a negative contribution to the next layer \citep{Nair2010}. The ReLU function helps solve the ``exploding/vanishing gradient'' problem because it is a non-saturated activation function. Saturated activation functions (e.g. sigmoid, tanh) have an undesirable property of producing a derivative close to zero when the activation saturates either tail of 0 or 1. This small gradient will reduce information from being propagated back to the previous layers' weights and prevent the network from learning efficiently \citep{Krizhevsky2012}. Additionally, we initialize the weights following a method in \cite{He2015} found to help networks (e.g. 30 convolutional/fully connected layers) converge and prevent saturation. The weight initialization draws random numbers from a Gaussian distribution centered on 0 with the standard deviation equal to $\sqrt{2/N}$ where $N$ is the number of input features.

	The advantage of a neural network is that it can be trained to identify subtle features inherent in a large data set. This learning capability is accomplished by allowing the weights and biases to vary in such a way as to minimize the difference (i.e. the cross-entropy) between the output of the neural network and the expected or desired value from the training data. The cross-entropy is used in our classifier over the mean squared error because it better represents the error in a binary classifier (transit vs. non-detection). We are interested in separating the input data by a line (or plane) which allow us to make predictions based on whether the input data falls above or below the dissecting surface. Suppose we have a model that predicts 2 classes with a ground truth (correct) label as $y$ and predicted probability $\hat{y}$ which is from the output of the neural network. The probability of predicting the correct value from our data for a given sample based on our two classes would be

\begin{equation} \label{model_prob}
  P(\hat{y}|X) = \hat{y}^{y} (1-\hat{y})^{(1-y)}
\end{equation}

\noindent Taking the logarithm and changing sign yields

\begin{equation} \label{model_prob_log}
  -\log P(\hat{y}|X) = -y\log{\hat{y}} -(1-y)\log{(1-\hat{y})}
\end{equation}

\noindent Summing Equation \ref{model_prob_log} over all our samples (i.e. training samples) and then dividing by the number of samples, $M$, yields the cross-entropy for our system,

\begin{equation} \label{cross_entropy}
 Loss = -\frac{1}{M} \sum_{i}^{M} y_{i}\log{\hat{y}_{i}} + (1-y_{i})\log{(1-\hat{y}_{i})}
\end{equation}

In effect we are minimizing the {\it Loss Function}, which tells us how well we are achieving our goal. The cross entropy is related to the expectation of the logarithmic difference between predictions and true labels where the expectation is taken using the true probabilities. As the $Loss$ goes to 0, the closer we are to predicting the true labels of the data, $y$. 

\begin{figure}
  \begin{center}
  \includegraphics[scale=0.6]{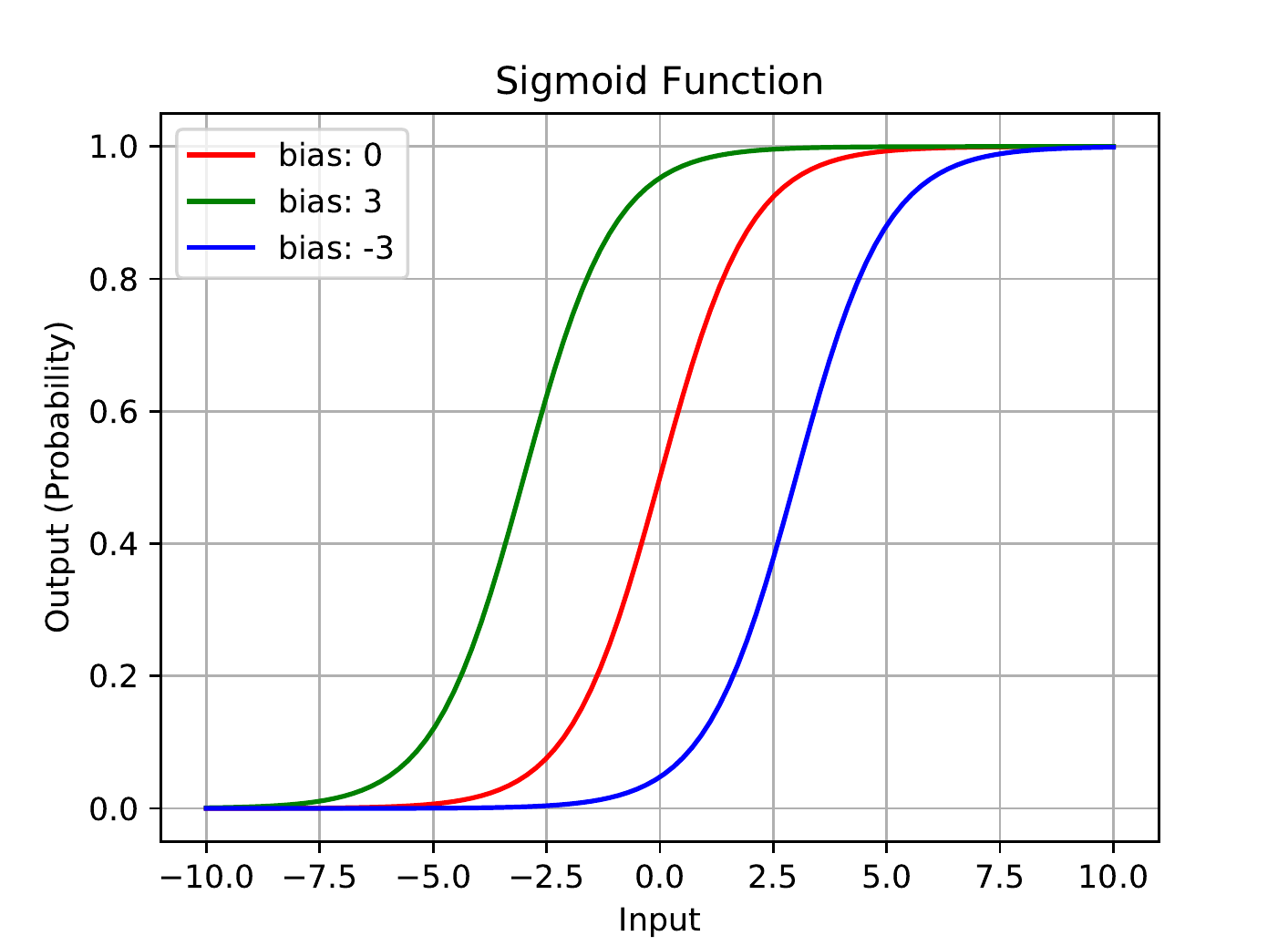}
  \end{center}
  \caption{ A sigmoid function (see equation \ref{neuron_last}) is used to compute the output probability for the decision of our neural network. The bias term shifts the output activation function to the left or right, which is critical for successful learning.  }
\label{sigmoid}
\end{figure}

The last layer in our network is made of only one neuron that is slightly different than the previous layer neurons because it classifies the input data into a transit or non-transit detection. The output of the last layer is modified with a sigmoid function (see equation \ref{neuron_last} and Figure \ref{sigmoid}). The sigmoid function is an ``activation function'', $f$, for the last neuron only and it will output a probability that the input falls into one of the two classes (transit or not). The input into the sigmoid function is the weighted sum of the input parameters (i.e. the output of the previous layer's neurons). 

\begin{equation} \label{neuron_last}
\begin{split}
  \text{\normalfont\ Output Activation: } \quad & f( \textbf{w} \cdot \textbf{X} + b) = \frac{1}{1+e^{-( \textbf{w} \cdot \textbf{X} + b)}} \\
\end{split}
\end{equation}

\noindent Optimization of the weights, $w_{j}$, for our deep net is done by minimizing the loss function of our system, the cross-entropy. The weights in each layer are optimized using a backward propagation scheme \citep{Werbos1974}. The weights are randomly initialized and after the loss function is computed, a backward pass propagates it from the output layer to the previous layers providing each weight parameter with an update value meant to decrease the loss. The new weight value is computed using the method of stochastic gradient descent such that

\begin{equation} \label{grad_descent}
 W^{i+1} = W^{i} - \epsilon ( \nabla Loss^{i}_{W} + \eta \nabla Loss^{i+1}_{W} )
\end{equation}

where $i$ is the iteration step, $\eta$ is the momentum, $\epsilon$ is the learning rate with a value larger than 0 and $\nabla$$Loss^{i}_{W}$ is the gradient of the loss function at the current iteration with respect to the current weights. We employ the use of Nesterov momentum to modify the weight update by predicting the gradient at a new position and correcting the gradient at the current position \citep{nesterov1983}. We use the common technique of stochastic gradient descent (SGD), \citep{Bottou1991} whereby we determine the gradient of the loss function using subsets of the training data (here 128 samples). Training on batches of data is more efficient for weight optimization if the total number of training samples is much larger than the batch size. Additionally, when we train, we cycle through all of the training data 30 times but at each epoch the samples in the batches are randomized. We employ the use of dropout with 25$\%$ of our neurons on the first layer of each network. While training, dropout helps prevent the network from over-fitting by randomly zeroing a neuron's output \citep{Srivastava2014}. The use of dropout is to first-order equivalent to an L2 regularizer because it adaptively distorts the neuron data to control over fitting \citep{Wager2013}.

The neural network relies on a handful of parameters that define the architecture (e.g. the hidden layer size and learning rate) which affect the performance. Tuning of these parameters was accomplished from a grid search where we trained over 1000 different neural works and chose the configuration that yielded the best performance.  We use a hidden layer size of 64, 32, 8, 1, where each is the number of neurons in a respectively deeper layer of the network. The optimal parameters are a regularization weight of 0, learning rate of 0.1, momentum of 0.25 and a decay rate of 0.0001. The decay rate corresponds to the learning rate decreasing by $ \epsilon^{i+1} = \epsilon^i  / (1. + decay *i) $, where i is the iteration step. We tried different regularization terms between 10 -- 10e-6 at factors of 10 and found they only decreased our network's performance. A smaller learning rate usually gave us a smaller accuracy and higher loss over the same number of training epochs. Our optimal deep net uses 13,937 trainable weights across 105 neurons with 2492 neural connections or synapses. We refer to this algorithm as MLP in Table \ref{tab:classifiers}.


\subsection{Wavelet}

Wavelets allow us to represent a signal as a series of components where we can discard the "least significant" pieces of that representation and keep the original signal largely intact. Training a neural network on all the wavelet components will allow it to learn the most significant pieces while ignoring lesser ones that do not define the signal (e.g. the noise). We design an another fully connected (fc) deep net using the same structure as above except the input data undergoes a wavelet transform. We compute a discrete wavelet transform using the second order Daubechies wavelet (i.e. four moments) on each whitened light curve \citep{Daubechies1992}. The approximate and first detailed coefficients are appended together and then used as the input for our learning algorithm, Wavelet MLP.

\subsection{Convolutional Neural Network}

The photometric measurements from a light curve are correlated to one another through time. We can make use of convolutions (conv) to compute local features from time-ordered input data. Convolutional neural networks (CNN 1D) can be thought of as generating new input data that has been convolved with a specific filter. The weights within each filter are optimized in a similar manner as a fully connected layer. Using fully connected layers where the every input feature has a dedicated weight per neuron can quickly grow the number of trainable parameters a model has. CNNs utilize convolutions and down sampling to compute local properties of the data when they are correlated to one another. The input data are discretely convolved with a filter or kernel as such

\begin{equation} \label{convolution}
 s(t) = \sum_{a}{x(t-a)w(a)}
\end{equation}

where the $t$ is the time index, $x$ is an input data array, $a$ loops over each element in the filter and $w$ are the weights in the filter. After the data has been convolved with a filter we down sample by averaging every three data points together to reduce the number of features for the next layer. We use an average pooling layer to help reduce the scatter from sources of noise. The average pooling layer mimics binning observational measurements in time. We also tested taking the maximum value within bins of 3 (max pooling layer) but it performed with less accuracy than an averaging layer. Planet finding techniques in the past have used convolutions via a matched filter approach however the filters are hand designed and only one is used. Our CNN 1D uses 4 filters each containing 6 weights that are optimized using the training data. After the input data are convolved and down sampled, we concatenate each light curve and use it as the input for a fully connected network with a layer size of 64,32,8,1. We use a ReLU activation function after convolving the data and before the average pooling layer. Additionally, 30 training epochs are used to teach the network. The training time per epoch for CNN 1D (our most expensive model) is 18 seconds on an Intel i7-7500U using TensorFlow \citep{tensorflow2015}. The code for each network is provided online\footnote{https://github.com/pearsonkyle/Exoplanet-Artificial-Intelligence}.

\begin{figure}
\centering
\hspace*{-.25in}
\includegraphics[scale=0.65]{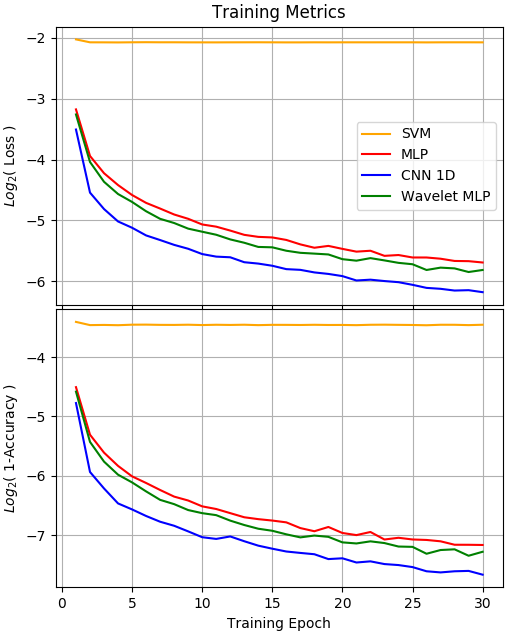}
\caption[ROC]{ The training performance of each algorithm is plotted as a function of training epoch. Each of our algorithms use the stochastic gradient descent (SGD) method to optimize the weights. The SGD solver uses a learning rate of 0.1, momentum of 0.25 and decay rate of 0.0001 with Nesterov momentum. The loss function for each model is the cross-entropy for a binary classifier (See equation \ref{cross_entropy}). Our SVM algorithm is over-fitting across 30 training epochs because the network's performance does not improve past 3 epochs. }
\label{training}
\end{figure}

\section{Cross-Validation and Algorithm Comparison}

Each models uses 311040 training samples in batches of 128 over 30 epochs to learn the transit features (See Figure \ref{training}). The validation/test data set is not used to train any model. The validation set consists of 933120 samples that span a larger range of noise than the training data (see \ref{tab:training}). 

\subsection{Multilayer Perceptron}
The large amount of weights in any deep net can create a non-convex loss function where there exists more than one local minimum. Therefore different random weight initializations can lead to different validation accuracies. We find variations in the accuracies of our deep learning algorithms on the order of $\sim$0.1$\%$ suggesting our use of a SGD solver is robust. 


\subsection{Support Vector Machine}
A SVM is a non-probabilistic linear classifier which constructs a hyperplane that optimally separates data points of two clusters given the labels \citep{Vapnik63}. A SVM is inherently designed to separate linear data but can be extended to separate non-linear data by transforming the input with a kernel (e.g. RBF). We use the method of SGD to optimize the weights from batches of 128 samples in our SVM. The method of SGD has been successfully applied to large-scale problems with more than 10$^{5}$ training samples \citep{Bottou2010}. A downside, similar to our MLP, is that the algorithm is sensitive to feature scaling so the data is whitened before training (mean=0 with unit variance). Based on a grid search of parameters we determined the optimal classifier has a cross-entropy loss function, linear kernel and an L2 regularization coefficient of 0.01. Regularization helps the weights from growing too large by introducing an additional term on the loss function that takes into account the size of each weight. 

\subsection{Least-Squares Box Fitting}

We compare our machine learning techniques to a more traditional model for detecting single transit events by fitting a box function to data (see Figure \ref{BLS}). However, using a step function at the boundary of the transit does not allow for sub-pixel precision when finding the duration or mid transit because a step function is not continuous. This presents a problem because the optimization function (i.e. $\chi^{2}$) has a Jacobian with zero at multiple locations suggesting the current parameters require no updating and thus convergence has been reached. Instead, we use a 1 pixel slope to define the boundary of our transit such that the depth at the 1 pixel slope is half that of the full transit depth. Adding a slope allows us to evaluate our transit function with sub-pixel precision because it is a continuous function. We optimize our parameter estimation using a simplex algorithm, Nelder-Mead, and minimize the mean squared difference between the data and model \citep{Nelder1965}. Random parameter initializations often lead least-squares optimizations away from the global solution within noisy data. The local solution will depend on the initial conditions and optimization algorithm. We were generous by initializing the BLS routine with a mid transit value 20 minutes different (10 data points) than the actual mid transit value and all other parameters fixed to the true value. The BLS algorithm is not a probabilistic classifier like the MLP or CNN so we create a score function such that it can be compared in the receiver operating characteristic (ROC) plot. The score of the transit fit is calculated as a combination of accuracies for mid transit and transit depth. The accuracy score makes sure the mid transit is within our time series and close to the center of the data and that the transit depth is greater than 0 and in most cases, greater than the noise.

\subsection{Algorithm Comparison}
\begin{figure}
\centering
\hspace*{-.25in}
\includegraphics[scale=0.65]{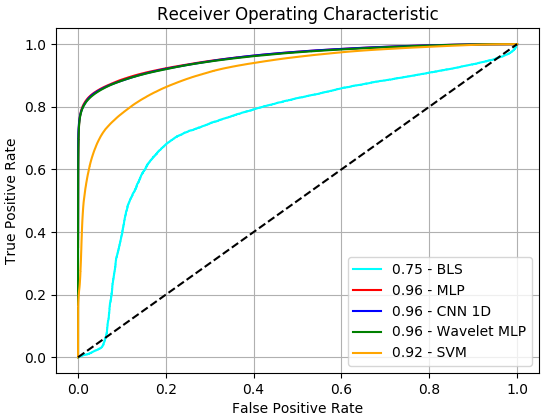}
\caption[ROC]{ We compare the classification of different transit detection algorithms; a fully connected neural network (MLP), convolutional neural network (CNN), fully connected network with wavelet transformed input (Wavelet MLP), a support vector machine (SVM) and a box fitting Least Squares method (BLS). The ROC plot shows the cumulative distribution of true positives and false negatives as the discrimination threshold is varied. The discrimination threshold and the probability output from the deep nets are used to classify the input. BLS and SVM are non-probabilistic algorithms so we make up an artificial score based on the accuracy of the transit depth and mid transit time such that we can compare it to our other models. We use all of the test data to calculate the ROC plot and report the area under each curve in the legend. A perfect classifier has an area of 1.}
\label{roc}
\end{figure}

We use a receiver operating characteristic (ROC) plot to compare the results of each transit finding algorithm (See Figure \ref{roc}). The ROC plot illustrates the performance of a classifier as its discrimination threshold is varied (\citealt{Hanley1982}). A classifier outputs a probability for a sample pertaining to a specific class and it is often just rounded up to signify a transit or non-transit detection (0 or 1). The ROC plot changes the classification of each model by varying where the probability data gets rounded. The true positive rate (TPR) is known as the probability of detection and taking 1-TPR yields the false negative rate. The ROC plot shows the cumulative distribution of the TPR as the discrimination threshold is varied against the cumulative distribution of false positive rates (FPR). The true negative rate can be calculated as 1-FPR. Ideally, the area under the curve should be close to unity which indicates a perfect classifier. The ROC plot shows the accuracy of each algorithm applied to test data the algorithms have not seen before.

\begin{table*}
  \caption{Summary of Classifiers}
  \label{tab:classifiers}
  \begin{center}

    \begin{tabular}{lrrrrr}
                         & BLS    & SVM    & MLP      & CNN 1D  & Wavelet MLP \\
    \hline &&&&&\\
  Input features          & 180    & 180    & 180      & 180     & 182    \\
  Trainable Params        & 3      & 181    & 13,937   & 17,293  & 14,169 \\
  Layers                  &        & 1      & 4        & 5       &  4     \\
  Total Neurons           &        & 1      & 105      & 109     &  105   \\
  Neural Connections      &        & 1      & 2492     & 2544    &  2494   \\
  &&&&& \\
  Training Accuracy  (\%) & 73.51  & 91.08  & 99.72  & 99.60  & 99.77  \\
  Training False Pos.(\%) & 22.34  & 3.05   & 0.08   & 0.21   & 0.08 \\
  Training False Neg.(\%) &  4.10  & 5.85   & 0.20   & 0.19   & 0.15  \\
  &&&&&\\
  Sensitivity Test (\%)  & 63.14  & 83.10   & 88.73   & 88.45   & 91.50  \\
  Test False Pos. (\%)   & 31.58  & 2.92    & 0.29    & 0.25    & 0.29  \\
  Test False Neg. (\%)   & 5.37   & 13.98   & 10.97   & 11.29   & 11.22  \\

  \end{tabular}
  \end{center}
\end{table*}

\section{Planet Detection Sensitivity}

\begin{figure}
\centering
\hspace*{-.25in}
\includegraphics[scale=0.7]{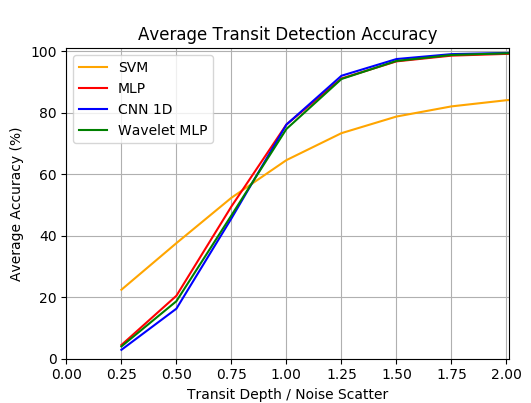}
\caption[Detection Accuracy]{ We test the sensitivity of our deep nets to data outside the range we trained it on. We generate 77760 light curves for each noise value. We find that the size of the transit depth does not influence the accuracy. Instead, the ratio of transit depth to noise dictates the accuracy of each detection algorithm. Based on this plot we can estimate the number of light curves required to significantly detect a planet below the noise by binning data together. }
\label{detection}
\end{figure}

We explore the sensitivity of our algorithms in order to understand the detection limits and robustness of finding new planets. We use our test data set to explore the accuracy under varying amounts of noise (see sensitivity test in Table \ref{tab:training}). Figure \ref{detection} shows the accuracy of our deep learning algorithms on data with varying amplitudes of noise. All of the initial parameters were kept the same as in Table \ref{tab:training} except for the noise parameter, this leads to 77760 samples in each noise bracket of Figure \ref{detection}. By whitening the data we remove the scale of the transit depth from the observations such that the limiting detection factor is the noise parameter. As the transit depth approaches the noise the accuracy of detection goes down. Figure \ref{detection} provides us with a means to estimate how much binning is required to reduce the scatter to be able to detect small planets.

\subsection{Time Series Variation}
\begin{figure}
\centering
\hspace*{-.25in}
\includegraphics[scale=0.64]{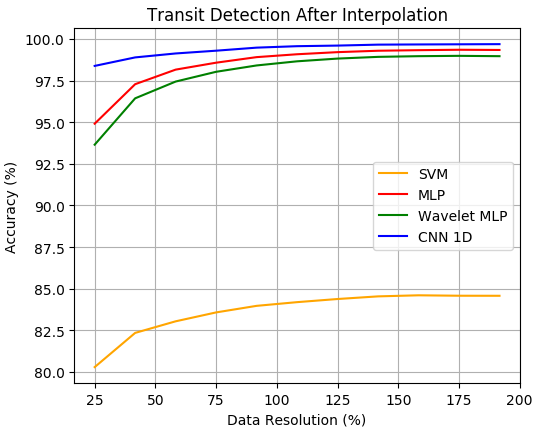}
\caption[FeatureLoss]{ We test our algorithms against new data using the parameters in Table 1 except the cadence or sampling rate is varied. This data resolution is compared to our original test data set and reported as a percentage on the x-axis (e.g. a 25$\%$ resolution corresponds to a cadence 4 times larger than the original). The new data has a different sampling rate for the same sized window in time but we linearly interpolate the data back onto the original grid size (i.e. 180 input features). The convolutional neural network has the best performance with lower resolution data because it pools local information together. CNN 1D could be used on observations with different sampling rates while suffering a minimal decrease in performance ($\sim$2$\%$) without training a new model on instrument specific properties. Data with a noise parameter above 1 was used in this analysis. }
\label{Interpolation}
\end{figure}

The nature of our deep net allows us to evaluate data from a different time series than the original training data using interpolation. The flexibility in the evaluation of different data stems from the networks ability to analyze 180 input features. Creating a new deep net for more instrument specific cadences is not necessary and we demonstrate this in Figure \ref{Interpolation}. \cite{Waldmann2016} argues that one can interpolate a lower cadence signal onto a higher cadence grid. The interpolation would incur an extra noise penalty but still provide enough information for his deep net to make a significant classification. We tested that hypothesis by creating a new data set of varying resolutions compared to the original data and tested the performance of each algorithm. Since the new data have a different sampling rate for the same sized window in time, we linearly interpolate the data back onto the original grid size (i.e. 180 input features). We find that the convolutional neural network has the smallest performance drop when evaluating down sampled data. The accuracy of detection remains the same  when evaluating data from a higher resolution grid. Arguably, data from unevenly sampled grids could also be interpolated onto a more uniform grid prior to input. 

Our deep nets were originally trained on data within a 6 hour time window at a 2 minute cadence with the transit duration being between at least 30 minutes to 4 hours. The deep net only knows about 180 input features and nothing regarding the time domain, except that the input is time ordered (an important property for the CNN 1D algorithm). This allows us to stretch the boundaries of our data interpretation to different time domains. We can use the same deep net to detect planets within the Kepler data even though it has a time cadence of 30 minutes. The longer cadence limits the time domain to a 90 hour window with detectable transits being between 7.5 hours to 60 hours. These conditions might be beneficial for some planets but it will not find those with shorter transit durations. Given our findings above, data with a lower cadence could be super-sampled and then evaluated with a minimal decrease in performance ($\sim$2$\%$). However, for the purposes of this pilot test we use the native resolution of the Kepler data during our analysis. 

\subsection{Feature Loss}
\begin{figure}
\centering
\hspace*{-.25in}
\includegraphics[scale=0.6]{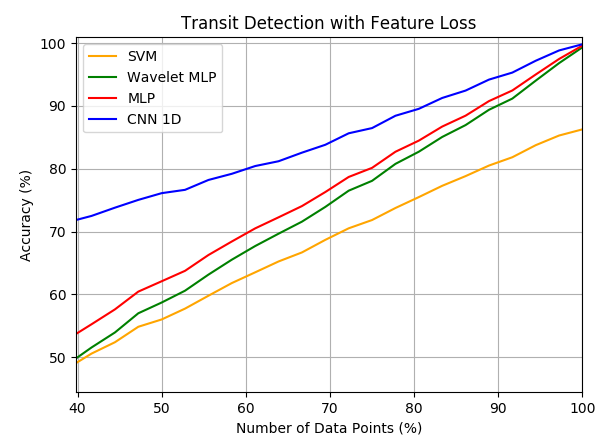}
\caption[FeatureLoss]{ We explore the capability of our deep nets to evaluate data with an incomplete set of measurements. We randomly remove chunks of data from each light curve up to certain extent. For each value, $i$, between 0 to 60$\%$ we select a random integer of chunks between 1 and $i$/10 to take out of the data. The positions of the chunks are randomized and chosen such that there are no overlaps which conserve the amount of data removed, $i$. The convolutional neural network has the best performance with less data because it pools local information together. Test data with a noise parameter above 1 was used in this analysis.}
\label{FeatureLoss}
\end{figure}

Observations are often subject to less than desirable conditions that suffer from instrumental malfunctions or weather induced systematics which can yield an incomplete set of measurements. We explore the capability of our deep learning algorithms to evaluate data that are missing features. Figure \ref{FeatureLoss} shows the evaluation of our algorithms with missing data points. We randomly remove chunks of data from each light curve up to a certain extent. For each value, $i$, between 0 to 60$\%$ we select a random integer of chunks between 1 and $i$/10 to take out of the data. The positions of the chunks are randomized and chosen such that there are no overlaps which conserve the amount of data removed, $i$. The convolutional neural network has the best performance with less data because it pools local information together. Test data with a noise parameter above 1 was used in this analysis.

\section{Time Series Evaluation}

\begin{figure*}
\label{timeseries}
\centering
\begin{adjustbox}{addcode={\begin{minipage}{\width}}{\caption{%
 We evaluate our neural network on time series data that spans 4 transits of an artificial planet with a period of 8.33 days. The artificial planet has a transit depth of 900 ppm where the noise parameter (($R_{p}^{2}$/R$_{s}^{2}$)/ $\sigma_{tol}$) is 0.87 after noise has been added. According to Figure \ref{detection} the CNN 1D will have a difficult time evaluating single transit detections at this noise level. However, binning the 4 transits together, where the dotted lines indicate the mid transit of each, we can detect the signature of the planet. The binned light curve has a noise parameter of 1.67 and a stronger planet detection rate than the individual transits. The opacity of the data points are mapped to the probability a transit lies within a subset of the time series. The period is determined using a brute force evaluation on numerous phase binned light curves and provides an initial constraint on the periodicity of the planet. The bottom subplots show a cumulative probability where the probabilities are summed up for each time window the algorithm evaluated. The whole time series was input into CNN 1D with overlapping windows every 5 data points.
      }\end{minipage}},rotate=0,center}
\includegraphics[scale=0.6]{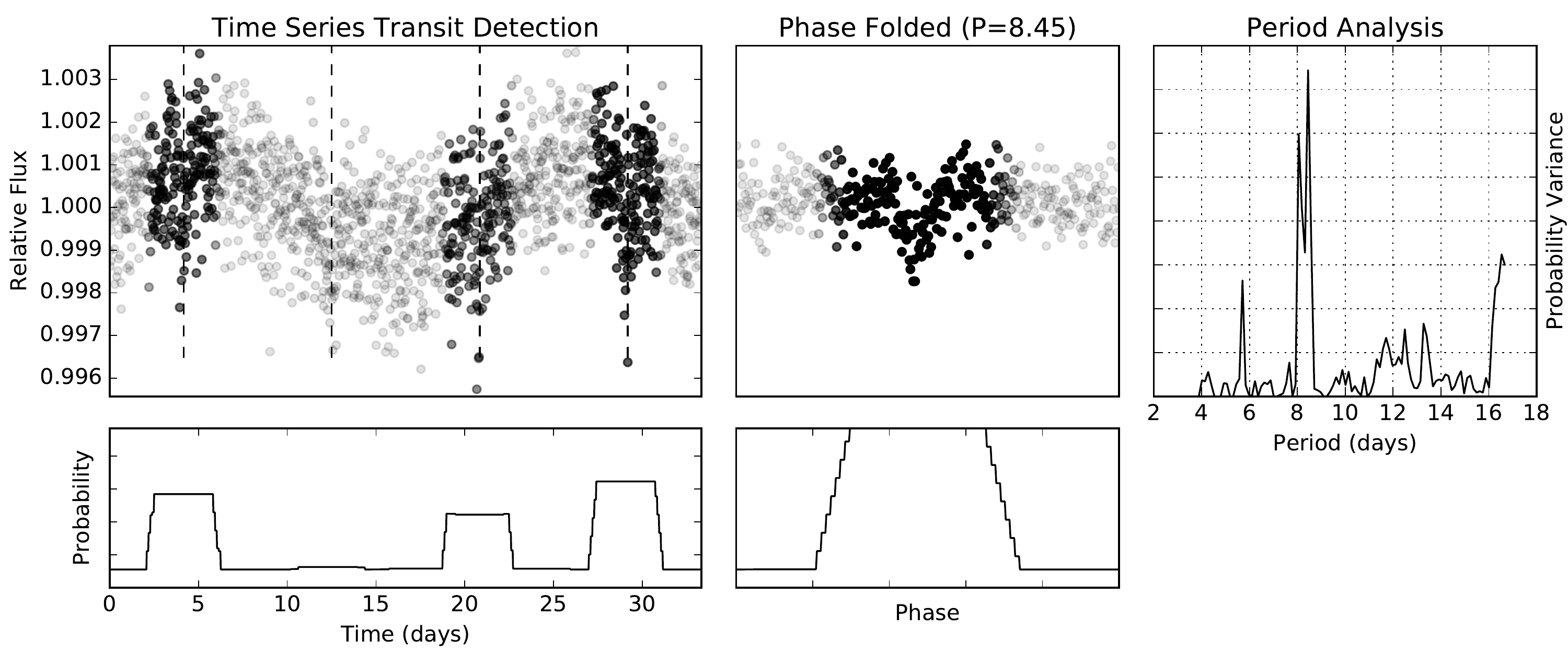}
  \end{adjustbox}
\label{timeseries}
\end{figure*}

Real mission data is seldom as ideal as the training set. To test the ability of our deep nets we apply them to longer time series data from Kepler. In order to evaluate data from a time series thats longer than the training data (i.e 180 input features) requires separating data into bite sized chunks for the neural network. We do this by creating overlapping patches from the time series that are offset by a small value. Algorithm 1 in the appendix highlights the specifics of performing our evaluation on longer time series data. After the input light curve is broken up, we compute a detection probability for each chunk and superimpose the probabilities in time. The cumulative probability time series or just probability times series, $PTS$, is normalized between 0 and 1 for the whole time series (See bottom left subplot in Figure 11). An estimate on the periodicity of the planetary transit can be made by finding the average difference in maxima within the $PTS$. Smoothing the $PTS$ plots using a Gaussian filter helps to find the maxima.


The search for ever smaller planets depends on beating down the noise enough to detect a signal. When an individual planet signal cannot be found in the data we use a brute force evaluation method for estimating periodic transits (Algorithm 2 in appendix). For a range of periods we compute the phase folded light curve and bin it to a specific cadence. In our test case, we simulate Kepler data and therefore bin the phase folded data to a 30 minute cadence. Next we follow the same steps as the time series evaluation above; break the data up into overlapping light curves and evaluate the cumulative probability phase series, $PPS$. However, we do this four different times for the same period but change the reference position/time for computing the phase. The second for loop in Algorithm 2 shows the step where we compute four different phase positions for the period. The phase curve is shifted to account for the possibility of the transit being at the edges of the data. Statistical fluctuations which might cause false positive detections are also mitigated with this cycling approach. Each phase curve is ran through Algorithm 1 and produces a $PPS$. The variance is computed at each phase value across all four $PPS$ arrays producing a probability variance phase series, $PVPS$. Each phase value in the $PVPS$ has a variance associated with it. The mean value of the $PVPS$ is then plotted with the corresponding period (e.g. see Period Analysis in the subplot of Figure 11).  

This method only works well for transits below the noise level. Larger transit signals will deconstructively add into the phase folded data but to an extent that can be greater than the scatter. Smaller transits do not have this problem and can stay hidden beneath the noise even when phase folded at the incorrect period. Only will the true signal reveal itself when the correct period is found and the signal is constructively added together to produce something above the noise. This method merely provides us with an orbital period estimate which we can then use as a prior for transit modeling routines.

\subsection{Kepler Data}

\begin{figure*}
\centering
\includegraphics[scale=0.65]{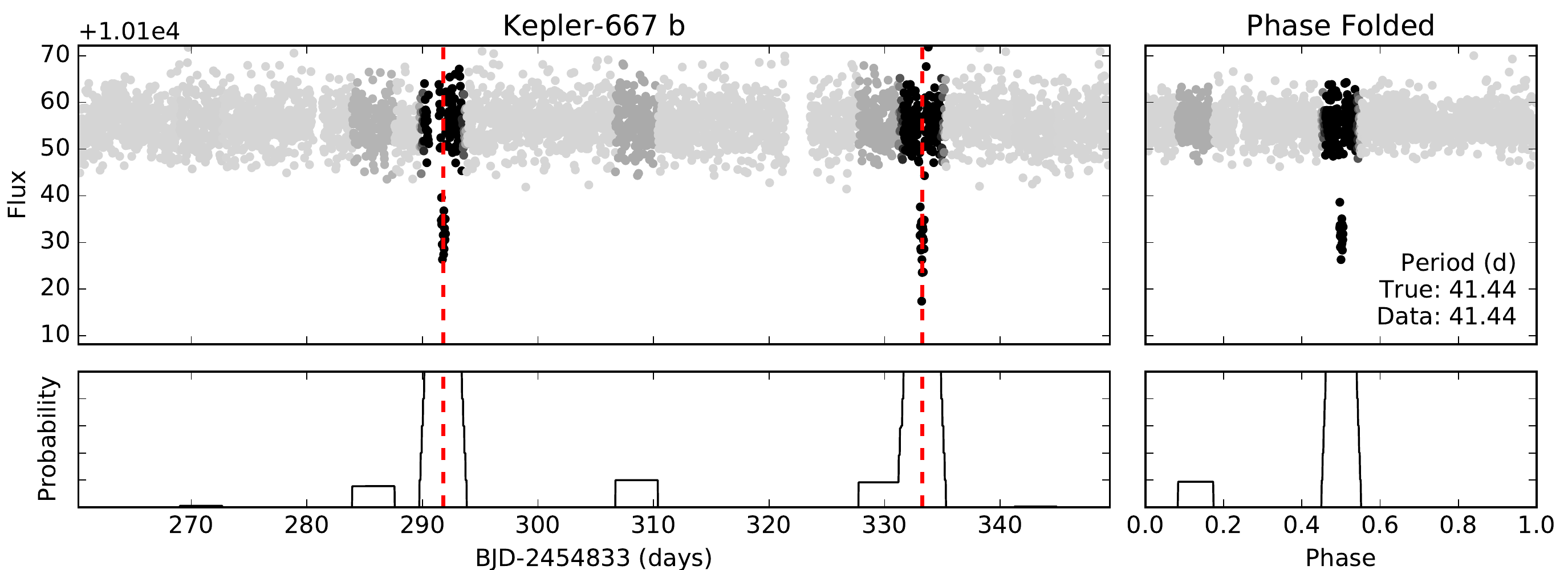}
\includegraphics[scale=0.65]{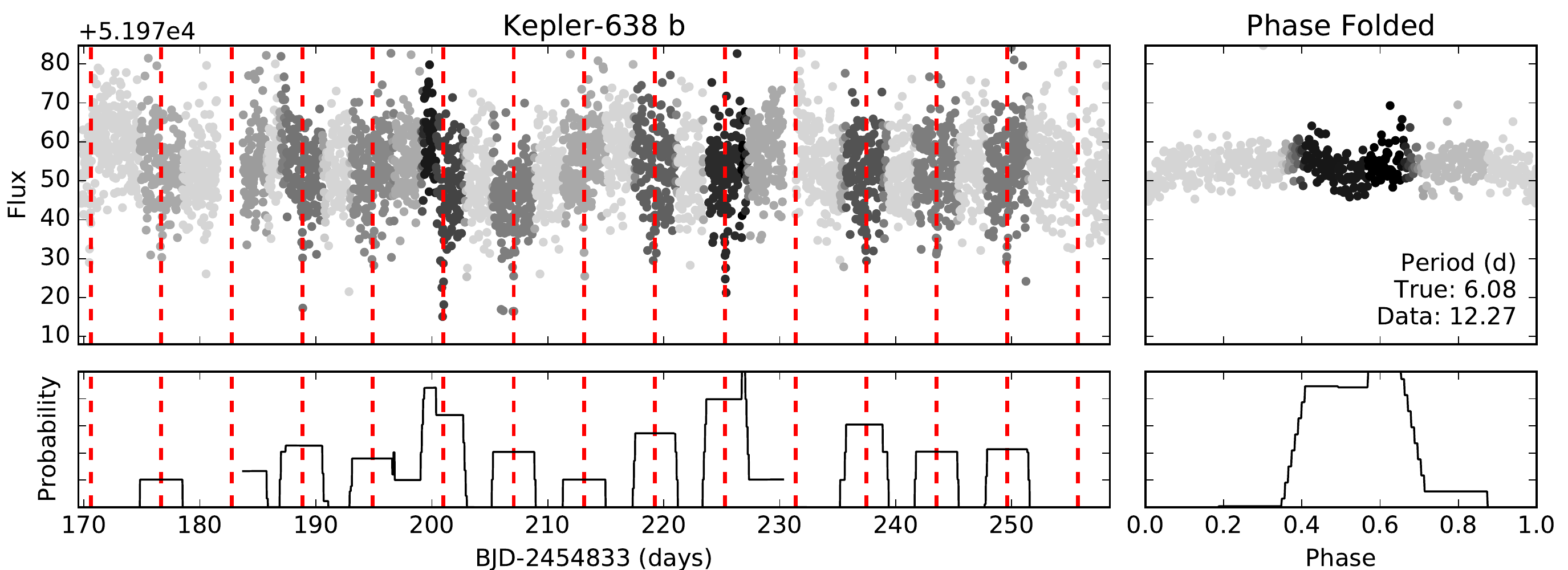}
\includegraphics[scale=0.65]{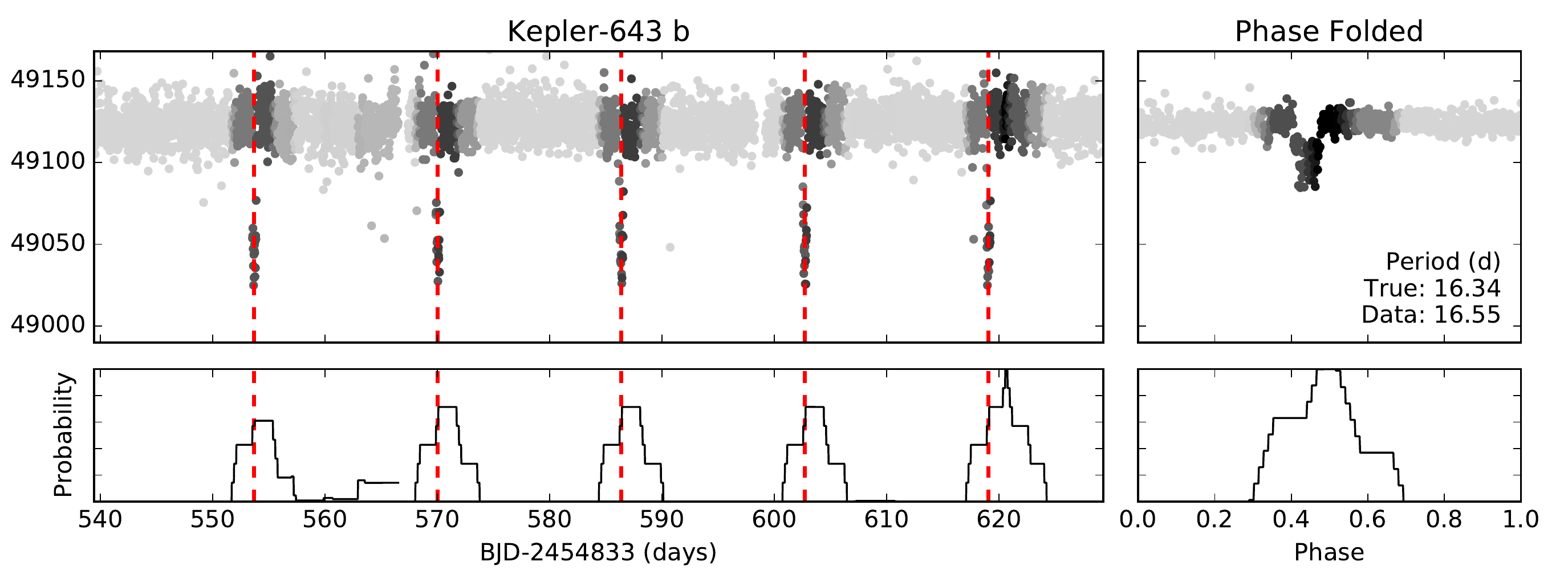}
 \caption[Time series]{ We evaluate a subset of the Kepler data set where the noise parameter is at least 2, the transit duration is between 7 and 15 hours and the period is less than 50 days. This limits the number of planets to have sufficient data in transit to make a robust single detection and have enough data to measure at least 2 transits. Each window of time is from a single Kepler quarter of data. The color of the data points are mapped to the probability of a transit being present. The red lines indicate the true ephemeris for the planet taken from the NASA Exoplanet Archive. The phase folded data is computed with a period estimated from the data. The period labeled ``Data'' is estimated by finding the average difference between peaks in the probability time plot (bottom left). This estimated period in most cases is similar to the true period and differs if the planet is in a multi-planet system or has data with strong systematics.   }
\label{KeplerTimeSeries1}
\end{figure*}

The Kepler exoplanet survey acquired over 3 years worth of data on transiting exoplanets that ranged in size from Earth to Jupiter-like. We only use one quarter of the Kepler data without any pre-processing to validate our neural network. Due to constraints from the time span the transit duration is limited to $\sim$7--15 hours with an orbital period greater than 90 hours. Detecting individual transit events with our algorithms requires a certain amount of features in-transit to yield an accurate prediction. We chose random targets with a noise parameter ($R_{p}^{2}$/R$_{s}^{2}$)/$\sigma_{tol}$) larger than 1.5 and a transit period greater than 90 hours so that we can acquire multiple transit events in a single quarter of Kepler data. Figure \ref{KeplerTimeSeries1} shows the results from a small analysis of Kepler targets. The probability plots are computed using the same method highlighted above for the time series evaluation. From the probability-time plot we can estimate the period of the planet by finding the average difference between peaks. The estimated orbital periods can deviate from the truth if discrepancies arise from multiple planet systems or strong systematics.


\section{Discussion}

Neural networks are an extremely versatile tool when it comes to pattern recognition. As demonstrated in \citet{Waldmann2016} deep nets can be trained to characterize planetary emission spectra as a means of narrowing the initial parameter space for atmospheric retrieval codes. Automated detection and characterization will pave the way for future planet finding surveys by eliminating human interaction which can bottleneck analyses and introduce error. 

Observations of exoplanets from different platforms contain separate observational limitations that can produce an incomplete set of measurements or sample the data differently. Accounting for each is best done using a convolutional neural network. Our CNN 1D algorithm achieved the highest performance on our interpolation and feature loss tests. The CCN 1D network computes local properties in the data using a filter convolution prior to input in a fully connected network. Additionally, the CNN 1D algorithms uses a down sampling technique by averaging which can help reduce some scatter in the data. Our findings indicate that data interpolated from a different resolution grid only suffered a 2$\%$ decrease in accuracy at most. The flexible nature of interpolation and network performance on different sized grids opens up the window of time for evaluating short and long period planets. 

\begin{figure}
\centering
\hspace*{-.25in}
\includegraphics[scale=0.65]{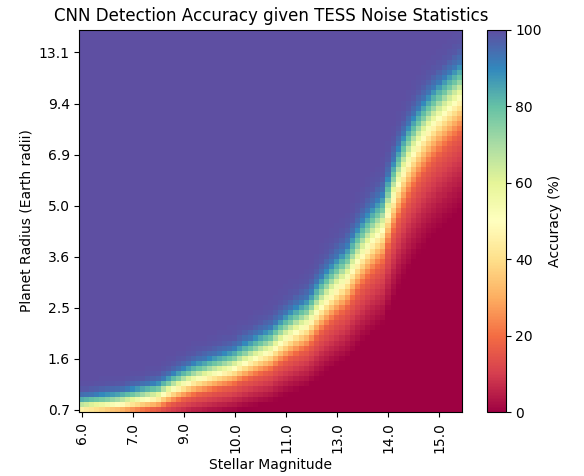}
\caption[FeatureLoss]{ We use the predicted photometric precision of TESS to explore the sensitivity of our deep net on detecting Earth-sized planets in the next generation planet survey. We assume a G type star with a radius equal to the sun when deriving the transit depth. Even though TESS will target mainly M dwarf stars, our values place a lower limit on transit depth and thus the detection accuracy. The CNN 1D algorithm can confidently detect a single Earth-like transit in bright stars (V<8) and will require binning the data to reach a greater SNR for dimmer stars.}
\label{TESS}
\end{figure}

The link between statistical properties of exoplanet populations and planetary formation models must account for the observational detection biases of exoplanet discovery. The transiting exoplanet survey satellite, TESS, will be flying next year and estimates of the planet discovery rate for our deep net are estimated in Figure \ref{TESS}. We use the average photometric precision of TESS shown in Figure 8 of \citet{Ricker2014} and our CNN 1D detection accuracy for various transit depths and noise levels. Figure \ref{TESS} shows the detection accuracy for potential planets assuming a solar type star with the same radius as the Sun. While TESS will target mostly M stars, which enhance the planet to star contrast, a solar type star places a lower limit on the transit depth. An Earth-sized exoplanet with a transit depth of 100 ppm around a 8th magnitude G star is 87$\%$ detectable with a noise parameter of 1.11. 

The BLS method for detecting single transit events without a priori knowledge of the transit location and stellar variability is not adequate at finding transits. The BLS algorithm on its own has trouble finding the correct solution because the optimization function has multiple minima. Different parameter initializations can guide the optimization algorithm into a local minimum. \citealt{Foreman-Mackey2015} find modeling the instrumental systematics along with the transit can greatly improve transit detection but at the cost of computational efficiency. \citealt{Foreman-Mackey2015} generate a synthetic data set similar to our own, with transit depths ranging between 400--10,000 ppm and inject noise on the order of 25--500 ppm based on Kepler Magnitude to photometric precision relations. This corresponds to a noise parameter,  $R_{p}^{2}$/R$_{s}^{2}$)/$\sigma_{tol}$, between 0.8 and 16 for the smallest transit depth tested in their sample. In all of our test cases the worst accuracy achieved with a noise parameter of 0.8 was $\sim$60\%. The worst detection accuracy in \citealt{Foreman-Mackey2015} was on the order of 1\% because shallower transits at longer periods were harder to detect for fainter stars. We predict our algorithm can detect small planets with large periods based on our findings from interpolating data from a super sampled grid. The average detection accuracy for a span of Kepler magnitudes in \citealt{Foreman-Mackey2015} was on the order of $\sim$60\% where we find our accuracies to be at least 60\% or above for noise parameters greater than 0.8. Figures 10, 11 and 12 from \citealt{Foreman-Mackey2015} were used to make this comparison. 

Performing a Monte Carlo simulation like a Markov chain or nested sampler can help find global solutions because they sample large portions of the parameter space. However, these algorithms require a long computational time and may not be a suitable means of analysis for big data sets. Current detection methods rely heavily on the uncertainty estimations of transit fitting algorithms when determining a significant signal. The uncertainty estimation is non-trivial but significant transits can be verified with three individual measurements to help decrease false positive signals.



\section{Conclusion}



In the era of ``big data'' manual interpretation of potential exoplanet candidates is a labor intensive effort and difficult to do with small transit signals (e.g. Earth-sized planets). Exoplanet transits have different shapes, as a result of, e.g. the stellar activity. Thus a simple template does not suffice to capture the subtle details, especially if the signal is below the noise or strong systematics are present. We use an artificial neural network to learn the photometric features of a transiting exoplanet. Deep machine learning is capable of processing millions of light curves in a matter of seconds. The discriminative nature of neural networks can only make a qualitative assessment of the candidate signal by indicating the likelihood of finding a transit within a subset of the time series. For planet signals smaller than the noise we devise a method for finding periodic transits using a phase folding technique that yields a constraint when fitting for the orbital period. Neural networks are highly generalizable allowing data to be evaluated with different sampling rates by interpolating the data to a standard grid. We validate our deep nets on light curves from the Kepler mission and detect periodic transits similar to the true period without any model fitting. Additionally, we test various methods to improve our planet detection rates including 1D convolutional networks and feature transformations such as wavelets and find significant improvement with CNNs. Machine learning techniques provide an artificially intelligent platform that can learn subtle features from large data sets in a more efficient manner than a human. 


The next generation of automation in data processing will be become more adaptive, self-learning and capable of optimizing itself with little to no auxiliary user input. The program will understand what it is looking at, make a qualitative pre-selection followed by a quantitative characterization of the exoplanet signal. In the future we would like to explore the use of more deep learning techniques (e.g. long short-term memory or PReLU) to increase the detection robustness to noise. Additionally, active research is currently being done in machine learning to optimize the network architecture and have it adapt to specific problems. Adding a pre-processing step has the potential to greatly improve the transit detection performance by removing systematics from the time series (e.g. \citealt{Aigrain2017}).

 A special thanks goes out to the anonymous reviewer for providing helpful comments and references, this manuscript would not be same without you.




\bibliographystyle{mnras}
\bibliography{ref} 

\begin{thebibliography}{}
\makeatletter
\relax
\def\mn@urlcharsother{\let\do\@makeother \do\$\do\&\do\#\do\^\do\_\do\%\do\~}
\def\mn@doi{\begingroup\mn@urlcharsother \@ifnextchar [ {\mn@doi@}
  {\mn@doi@[]}}
\def\mn@doi@[#1]#2{\def\@tempa{#1}\ifx\@tempa\@empty \href
  {http://dx.doi.org/#2} {doi:#2}\else \href {http://dx.doi.org/#2} {#1}\fi
  \endgroup}
\def\mn@eprint#1#2{\mn@eprint@#1:#2::\@nil}
\def\mn@eprint@arXiv#1{\href {http://arxiv.org/abs/#1} {{\tt arXiv:#1}}}
\def\mn@eprint@dblp#1{\href {http://dblp.uni-trier.de/rec/bibtex/#1.xml}
  {dblp:#1}}
\def\mn@eprint@#1:#2:#3:#4\@nil{\def\@tempa {#1}\def\@tempb {#2}\def\@tempc
  {#3}\ifx \@tempc \@empty \let \@tempc \@tempb \let \@tempb \@tempa \fi \ifx
  \@tempb \@empty \def\@tempb {arXiv}\fi \@ifundefined
  {mn@eprint@\@tempb}{\@tempb:\@tempc}{\expandafter \expandafter \csname
  mn@eprint@\@tempb\endcsname \expandafter{\@tempc}}}

\bibitem[\protect\citeauthoryear{Abadi et~al.,}{Abadi
  et~al.}{2015}]{tensorflow2015}
Abadi M.,  et~al., 2015, {TensorFlow}: Large-Scale Machine Learning on
  Heterogeneous Systems, \url {http://tensorflow.org/}

\bibitem[\protect\citeauthoryear{{Aigrain} et~al.,}{{Aigrain}
  et~al.}{2015}]{Aigrain2015}
{Aigrain} S.,  et~al., 2015, \mn@doi [\mnras] {10.1093/mnras/stv853}, \href
  {http://adsabs.harvard.edu/abs/2015MNRAS.450.3211A} {450, 3211}

\bibitem[\protect\citeauthoryear{{Aigrain}, {Parviainen}, {Roberts}, {Reece}
  \& {Evans}}{{Aigrain} et~al.}{2017}]{Aigrain2017}
{Aigrain} S.,  {Parviainen} H.,  {Roberts} S.,  {Reece} S.,   {Evans} T.,
  2017, preprint, \href {http://adsabs.harvard.edu/abs/2017arXiv170603064A} {}
  (\mn@eprint {arXiv} {1706.03064})

\bibitem[\protect\citeauthoryear{{Armstrong}, {Osborn}, {Brown}, {Kirk}, {Lam},
  {Pollacco}, {Spake}  \& {Walker}}{{Armstrong} et~al.}{2014}]{Armstrong2014}
{Armstrong} D.~J.,  {Osborn} H.~P.,  {Brown} D.~J.~A.,  {Kirk} J.,  {Lam}
  K.~W.~F.,  {Pollacco} D.~L.,  {Spake} J.,   {Walker} S.~R.,  2014, preprint,
  \href {http://adsabs.harvard.edu/abs/2014arXiv1411.6830A} {} (\mn@eprint
  {arXiv} {1411.6830})

\bibitem[\protect\citeauthoryear{{Armstrong}, {Pollacco}  \&
  {Santerne}}{{Armstrong} et~al.}{2016}]{Armstrong2016}
{Armstrong} D.~J.,  {Pollacco} D.,   {Santerne} A.,  2016, preprint, \href
  {http://adsabs.harvard.edu/abs/2016arXiv161101968A} {} (\mn@eprint {arXiv}
  {1611.01968})

\bibitem[\protect\citeauthoryear{{Auvergne} et~al.,}{{Auvergne}
  et~al.}{2009}]{Auvergne2009}
{Auvergne} M.,  et~al., 2009, \mn@doi [\aap] {10.1051/0004-6361/200810860},
  \href {http://adsabs.harvard.edu/abs/2009A%26A...506..411A} {506, 411}

\bibitem[\protect\citeauthoryear{{Bakos}, {Noyes}, {Kov{\'a}cs}, {Stanek},
  {Sasselov}  \& {Domsa}}{{Bakos} et~al.}{2004}]{Bakos2004}
{Bakos} G.,  {Noyes} R.~W.,  {Kov{\'a}cs} G.,  {Stanek} K.~Z.,  {Sasselov}
  D.~D.,   {Domsa} I.,  2004, \mn@doi [\pasp] {10.1086/382735}, \href
  {http://adsabs.harvard.edu/abs/2004PASP..116..266B} {116, 266}

\bibitem[\protect\citeauthoryear{{Borucki} et~al.,}{{Borucki}
  et~al.}{2010}]{Borucki2010}
{Borucki} W.~J.,  et~al., 2010, \mn@doi [Science] {10.1126/science.1185402},
  \href {http://adsabs.harvard.edu/abs/2010Sci...327..977B} {327, 977}

\bibitem[\protect\citeauthoryear{Bottou}{Bottou}{1991}]{Bottou1991}
Bottou L.,  1991, in Proceedings of Neuro-N\^imes 91. EC2, Nimes, France, \url
  {http://leon.bottou.org/papers/bottou-91c}

\bibitem[\protect\citeauthoryear{Bottou}{Bottou}{2010}]{Bottou2010}
Bottou L.,  2010, in Lechevallier Y.,  Saporta G.,  eds, Proceedings of the
  19th International Conference on Computational Statistics (COMPSTAT'2010).
  Springer, Paris, France, pp 177--187, \url
  {http://leon.bottou.org/papers/bottou-2010}

\bibitem[\protect\citeauthoryear{{Burke} et~al.,}{{Burke}
  et~al.}{2014}]{Burke2014}
{Burke} C.~J.,  et~al., 2014, \mn@doi [\apjs] {10.1088/0067-0049/210/2/19},
  \href {http://adsabs.harvard.edu/abs/2014ApJS..210...19B} {210, 19}

\bibitem[\protect\citeauthoryear{{Carpano}, {Aigrain}  \& {Favata}}{{Carpano}
  et~al.}{2003}]{Carpano2003}
{Carpano} S.,  {Aigrain} S.,   {Favata} F.,  2003, \mn@doi [\aap]
  {10.1051/0004-6361:20030093}, \href
  {http://adsabs.harvard.edu/abs/2003A%26A...401..743C} {401, 743}

\bibitem[\protect\citeauthoryear{{Carter} \& {Winn}}{{Carter} \&
  {Winn}}{2009}]{Carter2009}
{Carter} J.~A.,  {Winn} J.~N.,  2009, \mn@doi [\apj]
  {10.1088/0004-637X/704/1/51}, \href
  {http://adsabs.harvard.edu/abs/2009ApJ...704...51C} {704, 51}

\bibitem[\protect\citeauthoryear{{Coughlin} et~al.,}{{Coughlin}
  et~al.}{2016}]{Coughlin2016}
{Coughlin} J.~L.,  et~al., 2016, \mn@doi [\apjs] {10.3847/0067-0049/224/1/12},
  \href {http://adsabs.harvard.edu/abs/2016ApJS..224...12C} {224, 12}

\bibitem[\protect\citeauthoryear{{Crossfield} et~al.,}{{Crossfield}
  et~al.}{2015}]{Crossfield2015}
{Crossfield} I.~J.~M.,  et~al., 2015, \mn@doi [\apj]
  {10.1088/0004-637X/804/1/10}, \href
  {http://adsabs.harvard.edu/abs/2015ApJ...804...10C} {804, 10}

\bibitem[\protect\citeauthoryear{Daubechies}{Daubechies}{1992}]{Daubechies1992}
Daubechies I.,  1992, Ten Lectures on Wavelets.
Society for Industrial and Applied Mathematics, Philadelphia, PA, USA

\bibitem[\protect\citeauthoryear{{Foreman-Mackey}, {Montet}, {Hogg}, {Morton},
  {Wang}  \& {Sch{\"o}lkopf}}{{Foreman-Mackey}
  et~al.}{2015}]{Foreman-Mackey2015}
{Foreman-Mackey} D.,  {Montet} B.~T.,  {Hogg} D.~W.,  {Morton} T.~D.,  {Wang}
  D.,   {Sch{\"o}lkopf} B.,  2015, \mn@doi [\apj]
  {10.1088/0004-637X/806/2/215}, \href
  {http://adsabs.harvard.edu/abs/2015ApJ...806..215F} {806, 215}

\bibitem[\protect\citeauthoryear{{Fressin} et~al.,}{{Fressin}
  et~al.}{2013}]{Fressin2013}
{Fressin} F.,  et~al., 2013, \mn@doi [\apj] {10.1088/0004-637X/766/2/81}, \href
  {http://adsabs.harvard.edu/abs/2013ApJ...766...81F} {766, 81}

\bibitem[\protect\citeauthoryear{{George} \& {Huerta}}{{George} \&
  {Huerta}}{2017}]{George2017}
{George} D.,  {Huerta} E.~A.,  2017, preprint, \href
  {http://adsabs.harvard.edu/abs/2017arXiv170100008G} {} (\mn@eprint {arXiv}
  {1701.00008})

\bibitem[\protect\citeauthoryear{{Gibson}, {Aigrain}, {Roberts}, {Evans},
  {Osborne}  \& {Pont}}{{Gibson} et~al.}{2012}]{Gibson2012}
{Gibson} N.~P.,  {Aigrain} S.,  {Roberts} S.,  {Evans} T.~M.,  {Osborne} M.,
  {Pont} F.,  2012, \mn@doi [\mnras] {10.1111/j.1365-2966.2011.19915.x}, \href
  {http://adsabs.harvard.edu/abs/2012MNRAS.419.2683G} {419, 2683}

\bibitem[\protect\citeauthoryear{Hanley \& McNeil}{Hanley \&
  McNeil}{1982}]{Hanley1982}
Hanley J.~A.,  McNeil B.~J.,  1982, \mn@doi [Radiology]
  {10.1148/radiology.143.1.7063747}, 143, 29

\bibitem[\protect\citeauthoryear{{He}, {Zhang}, {Ren}  \& {Sun}}{{He}
  et~al.}{2015}]{He2015}
{He} K.,  {Zhang} X.,  {Ren} S.,   {Sun} J.,  2015, preprint, \href
  {http://adsabs.harvard.edu/abs/2015arXiv150201852H} {} (\mn@eprint {arXiv}
  {1502.01852})

\bibitem[\protect\citeauthoryear{{Howell} et~al.,}{{Howell}
  et~al.}{2014}]{Howell2014}
{Howell} S.~B.,  et~al., 2014, \mn@doi [\pasp] {10.1086/676406}, \href
  {http://adsabs.harvard.edu/abs/2014PASP..126..398H} {126, 398}

\bibitem[\protect\citeauthoryear{{Ioffe} \& {Szegedy}}{{Ioffe} \&
  {Szegedy}}{2015}]{Ioffe2015}
{Ioffe} S.,  {Szegedy} C.,  2015, preprint, \href
  {http://adsabs.harvard.edu/abs/2015arXiv150203167I} {} (\mn@eprint {arXiv}
  {1502.03167})

\bibitem[\protect\citeauthoryear{{Jenkins}, {Caldwell}  \& {Borucki}}{{Jenkins}
  et~al.}{2002}]{Jenkins2002}
{Jenkins} J.~M.,  {Caldwell} D.~A.,   {Borucki} W.~J.,  2002, \mn@doi [\apj]
  {10.1086/324143}, \href {http://adsabs.harvard.edu/abs/2002ApJ...564..495J}
  {564, 495}

\bibitem[\protect\citeauthoryear{{Kipping} \& {Lam}}{{Kipping} \&
  {Lam}}{2017}]{Kipping2017}
{Kipping} D.~M.,  {Lam} C.,  2017, \mn@doi [\mnras] {10.1093/mnras/stw2974},
  \href {http://adsabs.harvard.edu/abs/2017MNRAS.465.3495K} {465, 3495}

\bibitem[\protect\citeauthoryear{{Kov{\'a}cs}, {Zucker}  \&
  {Mazeh}}{{Kov{\'a}cs} et~al.}{2002}]{Kovacs2002}
{Kov{\'a}cs} G.,  {Zucker} S.,   {Mazeh} T.,  2002, \mn@doi [\aap]
  {10.1051/0004-6361:20020802}, \href
  {http://adsabs.harvard.edu/abs/2002A%26A...391..369K} {391, 369}

\bibitem[\protect\citeauthoryear{Krizhevsky, Sutskever  \& Hinton}{Krizhevsky
  et~al.}{2012}]{Krizhevsky2012}
Krizhevsky A.,  Sutskever I.,   Hinton G.~E.,  2012, in Pereira F.,  Burges C.
  J.~C.,  Bottou L.,   Weinberger K.~Q.,  eds, , Advances in Neural Information
  Processing Systems 25.
Curran Associates, Inc., pp 1097--1105, \url
  {http://papers.nips.cc/paper/4824-imagenet-classification-with-deep-convolutional-neural-networks.pdf}

\bibitem[\protect\citeauthoryear{{LSST Science Collaboration} et~al.,}{{LSST
  Science Collaboration} et~al.}{2009}]{LSST2009}
{LSST Science Collaboration} et~al., 2009, preprint, \href
  {http://adsabs.harvard.edu/abs/2009arXiv0912.0201L} {} (\mn@eprint {arXiv}
  {0912.0201})

\bibitem[\protect\citeauthoryear{{Mandel} \& {Agol}}{{Mandel} \&
  {Agol}}{2002}]{Mandel2002}
{Mandel} K.,  {Agol} E.,  2002, \mn@doi [\apjl] {10.1086/345520}, \href
  {http://adsabs.harvard.edu/abs/2002ApJ...580L.171M} {580, L171}

\bibitem[\protect\citeauthoryear{{McCauliff} et~al.,}{{McCauliff}
  et~al.}{2015}]{McCauliff2015}
{McCauliff} S.~D.,  et~al., 2015, \mn@doi [\apj] {10.1088/0004-637X/806/1/6},
  \href {http://adsabs.harvard.edu/abs/2015ApJ...806....6M} {806, 6}

\bibitem[\protect\citeauthoryear{{McQuillan}, {Mazeh}  \&
  {Aigrain}}{{McQuillan} et~al.}{2014}]{Mcquillan2014}
{McQuillan} A.,  {Mazeh} T.,   {Aigrain} S.,  2014, \mn@doi [\apjs]
  {10.1088/0067-0049/211/2/24}, \href
  {http://adsabs.harvard.edu/abs/2014ApJS..211...24M} {211, 24}

\bibitem[\protect\citeauthoryear{{Mislis}, {Bachelet}, {Alsubai}, {Bramich}  \&
  {Parley}}{{Mislis} et~al.}{2016}]{Mislis2016}
{Mislis} D.,  {Bachelet} E.,  {Alsubai} K.~A.,  {Bramich} D.~M.,   {Parley} N.,
   2016, \mn@doi [\mnras] {10.1093/mnras/stv2333}, \href
  {http://adsabs.harvard.edu/abs/2016MNRAS.455..626M} {455, 626}

\bibitem[\protect\citeauthoryear{{Morello}, {Waldmann}, {Tinetti}, {Howarth},
  {Micela}  \& {Allard}}{{Morello} et~al.}{2015}]{Morello2015}
{Morello} G.,  {Waldmann} I.~P.,  {Tinetti} G.,  {Howarth} I.~D.,  {Micela} G.,
    {Allard} F.,  2015, \mn@doi [\apj] {10.1088/0004-637X/802/2/117}, \href
  {http://adsabs.harvard.edu/abs/2015ApJ...802..117M} {802, 117}

\bibitem[\protect\citeauthoryear{Nair \& Hinton}{Nair \&
  Hinton}{2010}]{Nair2010}
Nair V.,  Hinton G.~E.,  2010, in Fürnkranz J.,  Joachims T.,  eds,
  Proceedings of the 27th International Conference on Machine Learning
  (ICML-10). Omnipress, pp 807--814, \url
  {http://www.icml2010.org/papers/432.pdf}

\bibitem[\protect\citeauthoryear{Nelder \& Mead}{Nelder \&
  Mead}{1965}]{Nelder1965}
Nelder J.~A.,  Mead R.,  1965, \mn@doi [The Computer Journal]
  {10.1093/comjnl/7.4.308}, 7, 308

\bibitem[\protect\citeauthoryear{Nesterov}{Nesterov}{1983}]{nesterov1983}
Nesterov Y.,  1983, in Soviet Mathematics Doklady. pp 372--376

\bibitem[\protect\citeauthoryear{Newell}{Newell}{1969}]{Newell1969}
Newell A.,  1969, \mn@doi [Science] {10.1126/science.165.3895.780}, 165, 780

\bibitem[\protect\citeauthoryear{{Pepper} et~al.,}{{Pepper}
  et~al.}{2007}]{Pepper2007}
{Pepper} J.,  et~al., 2007, \mn@doi [\pasp] {10.1086/521836}, \href
  {http://adsabs.harvard.edu/abs/2007PASP..119..923P} {119, 923}

\bibitem[\protect\citeauthoryear{{Petigura}, {Marcy}  \& {Howard}}{{Petigura}
  et~al.}{2013}]{Petigura2013}
{Petigura} E.~A.,  {Marcy} G.~W.,   {Howard} A.~W.,  2013, \mn@doi [\apj]
  {10.1088/0004-637X/770/1/69}, \href
  {http://adsabs.harvard.edu/abs/2013ApJ...770...69P} {770, 69}

\bibitem[\protect\citeauthoryear{{Pollacco} et~al.,}{{Pollacco}
  et~al.}{2006}]{Pollacco2006}
{Pollacco} D.~L.,  et~al., 2006, \mn@doi [\pasp] {10.1086/508556}, \href
  {http://adsabs.harvard.edu/abs/2006PASP..118.1407P} {118, 1407}

\bibitem[\protect\citeauthoryear{{Rauer} et~al.,}{{Rauer}
  et~al.}{2014}]{Rauer2014}
{Rauer} H.,  et~al., 2014, \mn@doi [Experimental Astronomy]
  {10.1007/s10686-014-9383-4}, \href
  {http://adsabs.harvard.edu/abs/2014ExA....38..249R} {38, 249}

\bibitem[\protect\citeauthoryear{{Ricker} et~al.,}{{Ricker}
  et~al.}{2014}]{Ricker2014}
{Ricker} G.~R.,  et~al., 2014, in Space Telescopes and Instrumentation 2014:
  Optical, Infrared, and Millimeter Wave. p. 914320 (\mn@eprint {arXiv}
  {1406.0151}), \mn@doi{10.1117/12.2063489}

\bibitem[\protect\citeauthoryear{Rosenblatt}{Rosenblatt}{1958}]{Rosenblatt1958}
Rosenblatt F.,  1958, Psychological Review, pp 65--386

\bibitem[\protect\citeauthoryear{{Rowe} et~al.,}{{Rowe}
  et~al.}{2015}]{Rowe2015}
{Rowe} J.~F.,  et~al., 2015, \mn@doi [\apjs] {10.1088/0067-0049/217/1/16},
  \href {http://adsabs.harvard.edu/abs/2015ApJS..217...16R} {217, 16}

\bibitem[\protect\citeauthoryear{Srivastava, Hinton, Krizhevsky, Sutskever  \&
  Salakhutdinov}{Srivastava et~al.}{2014}]{Srivastava2014}
Srivastava N.,  Hinton G.,  Krizhevsky A.,  Sutskever I.,   Salakhutdinov R.,
  2014, Journal of Machine Learning Research, 15, 1929

\bibitem[\protect\citeauthoryear{{Thompson}, {Mullally}, {Coughlin},
  {Christiansen}, {Henze}, {Haas}  \& {Burke}}{{Thompson}
  et~al.}{2015}]{Thompson2015}
{Thompson} S.~E.,  {Mullally} F.,  {Coughlin} J.,  {Christiansen} J.~L.,
  {Henze} C.~E.,  {Haas} M.~R.,   {Burke} C.~J.,  2015, \mn@doi [\apj]
  {10.1088/0004-637X/812/1/46}, \href
  {http://adsabs.harvard.edu/abs/2015ApJ...812...46T} {812, 46}

\bibitem[\protect\citeauthoryear{Vapnik \& Lerner}{Vapnik \&
  Lerner}{1963}]{Vapnik63}
Vapnik V.,  Lerner A.,  1963, Automation and Remote Control, 24

\bibitem[\protect\citeauthoryear{{Wager}, {Wang}  \& {Liang}}{{Wager}
  et~al.}{2013}]{Wager2013}
{Wager} S.,  {Wang} S.,   {Liang} P.,  2013, preprint, \href
  {http://adsabs.harvard.edu/abs/2013arXiv1307.1493W} {} (\mn@eprint {arXiv}
  {1307.1493})

\bibitem[\protect\citeauthoryear{{Waldmann}}{{Waldmann}}{2016}]{Waldmann2016}
{Waldmann} I.~P.,  2016, \mn@doi [\apj] {10.3847/0004-637X/820/2/107}, \href
  {http://adsabs.harvard.edu/abs/2016ApJ...820..107W} {820, 107}

\bibitem[\protect\citeauthoryear{{Waldmann}, {Tinetti}, {Deroo}, {Hollis},
  {Yurchenko}  \& {Tennyson}}{{Waldmann} et~al.}{2013}]{Waldmann2013}
{Waldmann} I.~P.,  {Tinetti} G.,  {Deroo} P.,  {Hollis} M.~D.~J.,  {Yurchenko}
  S.~N.,   {Tennyson} J.,  2013, \mn@doi [\apj] {10.1088/0004-637X/766/1/7},
  \href {http://adsabs.harvard.edu/abs/2013ApJ...766....7W} {766, 7}

\bibitem[\protect\citeauthoryear{Werbos}{Werbos}{1974}]{Werbos1974}
Werbos P.~J.,  1974, Beyond Regression: New Tools for Prediction and Analysis
  in the Behavioral Sciences.
IEEE Press

\bibitem[\protect\citeauthoryear{{Zellem}, {Griffith}, {Deroo}, {Swain}  \&
  {Waldmann}}{{Zellem} et~al.}{2014}]{Zellem2014}
{Zellem} R.~T.,  {Griffith} C.~A.,  {Deroo} P.,  {Swain} M.~R.,   {Waldmann}
  I.~P.,  2014, \mn@doi [\apj] {10.1088/0004-637X/796/1/48}, \href
  {http://adsabs.harvard.edu/abs/2014ApJ...796...48Z} {796, 48}

\makeatother
\end{thebibliography}





\appendix
\section{Algorithms for Time Series Evaluation}

\begin{algorithm}
\SetKwInOut{Input}{input}\SetKwInOut{Output}{output}

\Input{Light curve (more than 180 pts) \\ step size = 5 \\}

\Output{Probability time series}
\BlankLine
\emph{alloc probability time series, PTS, as array of zeros the same size as input light curve}\;
\BlankLine
 \For{$i:0$ \KwTo length(input)-180}{
	Prob = \texttt{Predict}( input[i:i+180] )\;
    PTS[i:i+180] += Prob\;
    \textit{increase i by step size}\;
 }
 \KwRet{PTS}\;

 \caption{Pseudo-code showing the steps we take to find transits on light curves with more than 180 data points. The algorithm evaluates overlapping bins in the time series in a boxcar-like manner. The colon in between two values represents all array values in between the first and second number. The \texttt{Predict} function returns a transit detection probability from our deep net, CNN 1D. We will refer to this algorithm as \texttt{TimeSeriesEval}. }
\end{algorithm}

\begin{algorithm}
\SetKwInOut{Input}{input}\SetKwInOut{Output}{output}

\Input{Light curve (more than 180 pts) \\ max Period, $P_{max}$ \\ min Period, $P_{min}$ \\ period step, $dP$ = 0.25}

\Output{Probability Variance per Period}
\BlankLine
\emph{alloc probability variance, PV, as array of zeros with ($P_{max}$ -- $P_{min}$ )/ $dP$ number of elements}\;
\BlankLine
 \For{P: $P_{min}$ \KwTo $P_{max}$}{
 
      \For{i: 0 \KwTo 0.8}{
        phase = (time-P*i)/P \% 1\;
        \emph{sort the phase values and input accordingly}\;
        \emph{bin the sorted phase series to specific cadence}\;
        PPS = \texttt{TimeSeriesEval}(\emph{sorted} phase series)\;
        \emph{save PPS to 2D array}\;
        \textit{increase i by 0.2}\;
        }
     
	\emph{Compute variance for each phase of 2D array producing a 1D array, PVPS}\;
    \emph{Save mean value of the PVPS}\;
    \textit{increase P by dP}\;
 }
 \KwRet{Mean Probability Variance of Each Period}\;

 \caption{ Pseudo-code showing the steps we take to evaluate phase folded data. The algorithm loops through a series of periods and computes the transit probability within overlapping bins of the phase series. The probability variance is computed to find deviations in the data representative of transit events. }
\end{algorithm}

\bsp	
\label{lastpage}
\end{document}